\documentclass[11pt]{article}
\usepackage{amsmath}
\usepackage{amssymb}     
\usepackage{graphicx}
\def \beq  {\begin{equation}}
\def \eeq  {\end{equation}}
\def \beqar {\begin{eqnarray}}
\def \eeqar {\end{eqnarray}}
\def \bed {{\begin{displaymath}}}
\def \eed {{\end{displaymath}}}
\def\sqr#1#2{{\vcenter{\vbox{\hrule height.#2pt
\hbox{\vrule width.#2pt height#1pt \kern#1pt
\vrule width.#2pt}\hrule height.#2pt}}}}

\def\la {{\langle}}
\def\ra {{\rangle}}
\def\Tr {{\rm Tr}}
\def\del {\partial}
\def\bdel{\bar{\partial}}

\def\vf {{\varphi}}

\def\bpi {{\bar \pi}}
\def\bPi {{\bar \Pi}}

\def\ep {\epsilon}

\def \dotA {{\dot A}}
\def \dotB {{\dot B}}

\def\bD {\bar{D}}
\def\bA {\bar{A}}

\def\bZ {{\bar Z}}

\def\bu {\bar{u}}

\def\bz {{\bar{z}}}

\def\D {{\mathcal D}}

\def\H {{\mathcal H}}

\def \CMP {{ Commun. Math. Phys.}}
\def \PRL {{ Phys. Rev. Lett.}}
\def \PL {{Phys. Lett. B}}
\def \NPBProc {{ Nucl. Phys. B (Proc. Suppl.)}}
\def \NP {{ Nucl. Phys.}}
\def \RMP {{ Rev. Mod. Phys.}}
\def \JGP {{ J. Geom. Phys.}}
\def \CQG {{ Class. Quant. Grav.}}
\def \MPL {{Mod. Phys. Lett.}}
\def \IJMP {{ Int. J. Mod. Phys.}}
\def \JHEP {{ JHEP}}
\def \PR {{Phys. Rev.}}
\def \JMP {{ J. Math. Phys.}}
\begin{document}

\begin{titlepage}
\null\vspace{-62pt} \pagestyle{empty}
\begin{center}
\rightline{CCNY-HEP-05/4}
\rightline{June 2005}
\vspace{1truein} {\Large\bf
Noncommutative mechanics, Landau levels, twistors}\\
~\\
{\Large\bf and Yang-Mills amplitudes}
\vskip .2in\noindent

\vspace{.6in}
V. P.~NAIR
\vskip .1in
{\it Physics Department\\
City College of the CUNY\\
New York, NY 10031}\\
E-mail: vpn@sci.ccny.cuny.edu\\
\vskip .1in

\vspace{.4in}
\centerline{\large\bf Abstract}
\end{center}
These lectures fall into two distinct, although tenouously related, parts. The first part is about fuzzy and
noncommutative spaces, and particle mechanics on such spaces, in other words,
noncommutative mechanics. The second part is a discussion/review of twistors
and how they can be used in the calculation of
Yang-Mills amplitudes. The point of connection between these two topics, discussed in the last section, is in the realization of holomorphic maps as the lowest Landau level wave functions, or as
wave functions of the Hilbert space used for the fuzzy version of the two-sphere.
This article is based
on lectures presented at the
conference on {\it Higher Dimensional Quantum Hall Effect and Noncommutative Geometry},
Trieste, March 2005, {\it Winter School on Modern Trends in Supersymmetric Mechanics},
Frascati, March 2005 and the {\it Montreal-Rochester-Syracuse-Toronto
Conference 2005}, Utica, May 2005.

\end{titlepage}

\pagestyle{plain} \setcounter{page}{2}

\section{Fuzzy spaces}
\label{sec:1}
\subsection{Definition and construction of ${\cal H}_N$}

Fuzzy spaces have been an area of research for a number of years by now
\cite{connes1}.
They have proved to be useful in some physical problems.
Part of the motivation for this
has been the discovery that noncommutative spaces, and more specifically
fuzzy spaces, can
arise as solutions in string and  $M$-theories \cite{mtheory}.
For example, in the matrix model version of
$M$-theory, noncommutative spaces can be
obtained as $(N\times N)$-matrix configurations
whose large $N$-limit will give smooth manifolds.
Fluctuations of branes are described
by gauge theories and, with this motivation,
there has  recently been  a large
number of papers dealing with gauge theories, and more
generally field theories, on such spaces \cite{douglas}.
There is also an earlier line of
development, motivated by quantum gravity, using the Dirac 
operator to
characterize the manifold and using `spectral actions' \cite{connes2}.

Even apart from their string and $M$-theory connections, fuzzy
spaces are interesting for other reasons. 
Because these spaces are
described by finite dimensional matrices, 
the number of possible modes for fields on such spaces is limited by
the Cayley-Hamilton theorem, and so, one has
a natural ultraviolet cutoff. We may think of such
field theories as a finite-mode approximation to 
commutative continuum field
theories, providing, in some sense, an alternative to lattice gauge
theories. Indeed, this point
of view has been pursued in some recent work \cite{fuzzfield}.

Analysis of fuzzy spaces and particle dynamics on such spaces 
are also closely related to the quantum Hall effect. 
The dynamics of charged particles
in a magnetic field can be restricted to the lowest 
Landau level, if the field is sufficiently strong, and 
this is equivalent to dynamics on a fuzzy version of the 
underlying spatial manifold. (The fact that the restriction to the
lowest Landau level gives noncommutativity of coordinates has been known for a long time;
for a recent review focusing on the fuzzy aspects, see \cite{KNR}.)

The main idea behind fuzzy spaces is the 
standard correspondence principle of the quantum theory, which is as
shown below.
\vskip .15in
\begin{center}
\begin{tabular}{l c l}
$\underline{Quantum~ Theory}$&~\hskip .5in$\hbar \rightarrow 0$\hskip .5in~ &$\underline{Classical~ Theory}$\\
&&\\
Hilbert space ${\mathcal H}$&&Phase space $M$\\
&$\longrightarrow$&\\
Operators on ${\mathcal H}$&& Functions on $M$\\
\end{tabular}
\end{center}
\vskip .1in\noindent
This correspondence suggests a new paradigm. Rather than dealing with theories on a continuous manifold
$M$, we take the Hilbert space ${\mathcal H}$ and the algebra of operators on it as the fundamental quantities and obtain the continuous manifold
$M$ as an approximation. Generally, instead of $\hbar$, we use an arbitrary deformation parameter $\theta$, so that the continuous manifold emerges not as the classical limit
in the physical sense, but as some other limit when $\theta \rightarrow 0$, which will mathematically mimic
the transition from quantum mechanics to classical mechanics.
The point of view where the spacetime manifold is not fundamental can be particularly satisfying in the context of quantum gravity, and in fact, it was in this context that the
first applications of noncommutative spaces to physics was initiated
\cite{connes2}.

Now passing to more specific details, by a fuzzy space, we mean 
a sequence 
$({\mathcal H}_N , Mat_N , {\cal D}_N )$,
where ${\mathcal H}_N$ is an $N$-dimensional Hilbert space,
$Mat_N$ is the matrix algebra of $N\times N$-matrices which act  
on
${\mathcal H}_N$, and ${\cal D}_N$ is a  
matrix
analog of the Dirac operator or, in many instances, just the matrix analog of the
Laplacian.
The inner product on the matrix algebra is given by
$\la A,B\ra = {1\over N} \Tr (A^\dagger B)$.
The Hilbert space  ${\mathcal H}_N$ leads to some smooth manifold
$M$ as $N\rightarrow \infty$. 
The matrix algebra $Mat_N$ approximates to the algebra of functions on
$M$. The operator
${\cal D}_N$ is needed to recover metrical and other geometrical
properties of the manifold
$M$. For example, information about the dimension of $M$ is contained in
the growth of the number of eigenvalues. More generally, noncommutative spaces are defined in a similar way, with a triple $({\mathcal H}, {\cal A}, {\cal D})$, where
${\mathcal H}$ can be infinite-dimensional, ${\cal A}$ is the algebra of operators
on ${\mathcal H}$ and ${\cal D}$ is a Dirac operator on ${\mathcal H}$
\cite{connes2}.
For fuzzy spaces the dimensionality of
${\mathcal H}$ is finite.

Rather than discuss generalities, we will consider the construction of some noncommutative and fuzzy spaces. Consider the flat $2k$-dimensional space
${\bf R}^{2k}$. We build up the coherent state representation buy considering the particle action \cite{geomquant}
\beq
{{\cal S}\over \theta} = -{i \over \theta} \int dt~  \bZ_\alpha {\dot Z}^\alpha
\label{fs1}
\eeq
where $\alpha = 1, 2, ..., k$. Evidently, the time-evolution of the variables
$Z^\alpha$, $\bZ_\alpha$ is trivial, and so, the theory is entirely
characterized by the phase space, or upon quantization, by
the specification of the Hilbert space.
From the action, we can identify the canonical commutation rules as
\beq
[ \bZ_\beta  ,Z^\alpha ] = \theta ~\delta^\alpha_\beta
\label{fs2}
\eeq
It is then possible to choose states, which are eigenstates of $Z^\alpha$,
defined by
$ \la z \vert Z^\alpha = \la z\vert z^\alpha$, so that wave functions
$f (z) = \la z \vert f\ra$ can be taken to be holomorphic.
The operators $Z^\alpha, ~\bZ_\beta$ are realized on these by
\beq
\begin{split}
Z^\alpha ~f(z) &= z^\alpha ~f(z)\\
\bZ_\beta ~f (z) &= \theta {\del \over \del z^\beta} ~f(z)\\
\end{split}
\label{fs3}
\eeq
The inner product for the wave functions should be of the form

\beq
\begin{split}
\la f \vert h\ra &= \int d\mu~ C(z, \bz ) ~ {\bar f} ~h\\
d\mu&= \prod_\alpha {dz^\alpha d\bz_\alpha \over (-2i)} \equiv \prod_\alpha d^2z_\alpha\\
\end{split}
\label{fs4}
\eeq
By imposing the adjointness condition $\la f \vert Z h \ra = \la \bZ f \vert h\ra$,
we get
\beq
\theta {\del C \over \del z^\alpha } = - \bz_\alpha C
\label{fs5}
\eeq
which can be solved to yield
\beq
\la f \vert h\ra = \int \prod_\alpha {d^2z_\alpha \over \pi \theta}
~ \exp \left[- {\bz_\alpha z^\alpha \over \theta}\right] ~ {\bar f} ~h\label{fs6}
\eeq
The overall normalization has been chosen so that the state $f=1$ has norm equal to $1$.

Since $f(z)$ is holomorphic, a basis of states can be given by $f= 1, ~z^\alpha,$
$ ~z^{\alpha_1} z^{\alpha_2},$ $ ...~$. The Hilbert space is infinite-dimensional and can be used for the noncommutative version of ${\bf R}^{2k}$.

\subsection{Star products}
The star product is very helpful in discussing the large $N$ limit.
(Star products have along history going back to Moyal and others.
The books and reviews quoted, \cite{connes1, douglas, geomquant} and others,
contain expositions of the star product.)  We shall
consider the two-dimensional case first, generalization to arbitrary even dimensions will be straightforward. A basis for the Hilbert space is given by
$1, ~ z, ~ z^2,$ etc., and using this basis, we can represent an operator
as a matrix $A_{mn}$. Associated to such a matrix, we define a function
$A(z, \bz )= (A)$, known as the symbol for $A$, by
\beqar
(A)= A(z, \bz )&\equiv& \sum_{mn} A_{mn}{ z^m~ \bz^n\over \sqrt{m!~ n!}} ~e^{-z\bz / \theta} \nonumber\\
&=& \sum_{mn} A_{mn}  \psi_m \psi^*_n\label{fs7}
\eeqar
where $\psi_n$ are given by
\beq
\psi_n = e^{- z \bz /2\theta}~{z^n \over \sqrt{n!}}
\label{fs8}
\eeq
These are normalized functions obeying the equation
\beq
\int {d^2z \over \theta\pi} ~ \psi^*_n ~\psi_m = \delta_{mn}
\label{fs9}
\eeq
The symbol corresponding to the product of two operators (or matrices) $A$ and $B$
may be written as
\beqar
(AB) &=& \sum \psi_m A_{mn} B_{nk} \psi^*_k\nonumber\\
&=& 
\sum \psi_m(z) A_{mn} \left[ \int {d^2w \over \theta\pi} \psi^*_n(z +w) \psi_r (z +w)
\right]  B_{rk} \psi^*_k (z)
\nonumber\\
&=& \int {d^2w \over \theta\pi}~ e^{-w{\bar w} /\theta} ~A(z, \bz +{\bar w}) ~B(z+w, \bz )
\label{fs10a}\\
&\equiv& (A)* (B)\nonumber\\
&=& (A) (B) ~+~ \theta {\del (A) \over \del \bz} {\del (B) \over \del z} ~+\cdots
\label{fs10b}
\eeqar
Functions on $M = {\bf C}$, under the star product, form an associative but noncommutative algebra. As $\theta$ becomes small, we may approximate the star product by the first two terms, giving
\beq
( ~[A, B]~) = (A)*(B) ~-~ (B)*(A) =
\theta \left( {\del (A) \over \del \bz} {\del (B) \over \del z} 
- {\del (B) \over \del \bz} {\del (A) \over \del z} \right)
\label{fs11}
\eeq
The right hand side is the Poisson bracket of $A$ and $B$, and this relation
is essentially the standard result that the commutators of operators tend 
to ($i$  times) the Poisson bracket of the corresponding functions 
(symbols) for small values of the
deformation parameter. In particular, we find
\beq
\begin{split}
Z * \bZ &= z\bz ~+~ \cdots\\
\bZ * Z &= \bz z +\theta +\cdots\\
Z * \bZ - \bZ * Z &= \theta ~+\cdots\\
\end{split}
\label{fs12}
\eeq
We can interpret $Z$, $\bZ$ as the coordinates of the space; they are noncommuting.
The noncommutativity is characterized by the parameter $\theta$, as we can use the
equations given above in terms of symbols to obtain the small $\theta$-limit.

These considerations can be generalized in an obvious way to $M= {\bf C}^k$.

\subsection{Complex projective space $CP^k$}

We shall now discuss the fuzzy version of
${\bf CP}^k$. Unlike the case of flat space, we will get a finite number of
states for ${\bf CP}^k$, say, $N$, so this will be a truly fuzzy space, rather than
just noncommutative. The continuous manifold ${\bf CP}^k$ can be obtained
as $N \rightarrow \infty$. In the previous discussion, we started with
continuous ${\bf R}^{2k}$, set up the quantum theory for the action
(\ref{fs1}), and the resulting Hilbert space could be interpreted as giving
the noncommutative version of ${\bf R}^{2k}$. We can follow the same strategy for
${\bf CP}^k$. In fact, we can adapt  the coherent state construction to obtain 
the fuzzy version of ${\bf CP}^k$. For a more detailed and group theoretic approach, see
\cite{KNR, geomquant, KN}.

Continuous ${\bf CP}^k$ is defined as the set of
$k+1$ complex variables $Z^\alpha$, with the identification of $Z^\alpha$ and 
$\lambda Z^\alpha$ where $\lambda$ is any nonzero complex number,
i.e., we start with ${\bf C}^{k+1}$ and make the identification
$Z^\alpha \sim \lambda Z^\alpha$, $\lambda \in {\bf C} -\{ 0\}$.
Based on the fact that there is natural action of $SU(k+1)$ on $Z^\alpha$ 
given by
\beq
Z^\alpha \longrightarrow Z'^\alpha = g^\alpha_{~\beta} ~Z^\beta, \hskip .5in g \in SU(k+1),
\label{fs13}
\eeq
we can show that ${\bf CP}^k$ can be obtained as the coset
\beq
{\bf CP}^k = {SU(k+1) \over U(k)}
\label{fs14}
\eeq
In fact, we may take (\ref{fs14}) as the definition of ${\bf CP}^k$.
The division by $U(k)$ suggests that we can obtain ${\bf CP}^k$ by considering a
``gauged'' version of the action (\ref{fs1}), where the gauge group is 
taken to be $U(k)$. Replacing the time-derivative by the covariant derivative,
the action becomes
\beq
{\cal S} = -i \int dt~ \bZ_\alpha ( \del_0 Z^\alpha -i A_0 Z^\alpha )
~-~ n \int dt~ A_0
\label{fs15}
\eeq
where we have also included a term for the gauge field.
(From now on, we will not display $\theta$ explicitly.)
This action is easily checked to be invariant under the $U(1)$ gauge transformation
\beq
Z^\alpha \rightarrow e^{i\vf } ~Z^\alpha , \hskip .7in A_0 \rightarrow A_0 +\del_0 \vf
\label{fs16}
\eeq
The pure gauge field part of the action is the one-dimensional Chern-Simons term.
The coefficient $n$ has to be quantized, following the usual arguments.
For example, we can consider the transformation where $\vf (t)$ obeys
$\vf (\infty ) - \vf (-\infty ) = 2\pi$. The action then changes by $-2\pi n$,
and since $\exp (i {\cal S})$ has to  be single-valued to have a well-defined
quantum theory, $n$ has to be an integer.

The variation of the action with respect to $A_0$ leads to the Gauss law for the theory,
\beq
\bZ_\alpha ~Z^\alpha -~ n ~\approx 0
\label{fs17}
\eeq
where the weak equality (denoted by $\approx$) indicates, as usual, that this conditon is to be imposed as a constraint.
This is a first class constraint in the Dirac sense, and hence it removes two 
degrees of freedom. Thus, from ${\bf C}^{k+1}$, we go to a space with $k$ complex dimensions. Given the $U(k)$ invariance, this can be identified as ${\bf CP}^k$.

The time-evolution of $Z^\alpha$ is again trivial and we are led to
the complete characterization of the theory by the Hilbert space, which must be obtained
taking account of the constraint (\ref{fs17}). In the quantum theory, the allowed physical states
must be annihilated  by the Gauss law.  Using the realization of the $Z^\alpha$,
$\bZ_\alpha$ given in (\ref{fs2}), this becomes
\beq
\left( z^\alpha {\del \over \del z^\alpha}  -~n \right) ~f(z) 
=0
\label{fs18}
\eeq
Thus the allowed functions $f(z)$ must have $n$ powers of $z$'s. They are of the form
\beq
f (z) = {1\over \sqrt{n!}}~ z^{\alpha_1}z^{\alpha_2}\cdots z^{\alpha_n}
\label{fs19}
\eeq
There are $N = (n+k)! / n! k! $ independent functions. The Hilbert space of such functions form the carrier space
of a completely symmetric rank $n$  irreducible representation of $SU(k+1)$.
A simple parametrization in terms off local coordinates on ${\bf CP}^k$ can be obtained
by writing
$ z^{k+1} = \lambda $,  $ z^i = \lambda \xi^i $, where $\xi^i = z^i /z^{k+1}
= z^i /\lambda$, for $i =1, 2, ..., k$. Correspondingly, the wave functions have the form
$f (z) = ~ \lambda^n ~f (\xi )$. The inner product for two such wave functions can be obtained from
the inner product (\ref{fs6}). We get
\beqar
\la f \vert h\ra &=& {1\over n!}~ \int {d^2\lambda \over \pi}
\prod_i {d^2\xi_i  \over \pi}
~e^{-\lambda {\bar\lambda}  (1 + {\bar\xi}\cdot \xi)}~ (\lambda {\bar\lambda})^{k+n}
~{\bar f}~ h \nonumber\\
&=& {(n+k)! \over n!~ k!}\int \left[{ k! \prod d^2\xi_i  \over \pi^k (1+{\bar\xi}\cdot \xi )^{k+1} }\right]~ {{\overline{ f(\xi )}}\over (1+{\bar\xi}\cdot \xi )^{n/2}}~ {h(\xi )\over (1+{\bar\xi}\cdot \xi )^{n/2}}
\nonumber\\
&=& N ~\int d\mu({\bf CP}^k)~ {{\overline {f(\xi )}}\over (1+{\bar\xi}\cdot \xi )^{n/2}}~ {h(\xi )\over (1+{\bar\xi}\cdot \xi )^{n/2}}  \label{fs20}
\eeqar
Here $N$ is the dimension of the Hilbert space and $d\mu ({\bf CP}^k)$ is the
standard volume element for ${\bf CP}^k$ in the local coordinates
$\xi^i$, ${\bar\xi}_i$.
Now, an $SU(k+1)$ matrix $g$ can be parametrized in such a way that
the last column $g^\alpha_{~ k+1}$ is given in terms of $\xi^i$, and
the factor $\sqrt{1+{\bar \xi}\cdot \xi}$, as
\beq
g = \left[ \begin{matrix} ~~.&~~.&~~.&~~.&~~\xi^1\\
~~.&~~.&~~.&~~.&~~\xi^2\\
~~.&~~.&~~.&~~.&~.\\
~~.&~~.&~~.&~~.&~~\xi^k\\
~~.&~~.&~~.&~~.&~~1\\
\end{matrix}\right]~{1\over \sqrt{1+{\bar\xi}\cdot \xi}}
\label{fs21}
\eeq
The states $f$ are thus of the form
\beq
{f(\xi )\over (1+{\bar \xi}\cdot \xi )^{n/2} } =
g^{\alpha_1}_{~k+1}g^{\alpha_2}_{~k+1}\cdots g^{\alpha_n}_{~k+1}
\label{fs22}
\eeq

Let $\vert n,r\ra$, $r = 1, 2, \cdots , N$, denote the states of the rank $n$ symmetric representation
of $SU(k+1)$. Then the Wigner $\D$-function corresponding to
$g$ in this representation is defined by $\D^{(n)}_{rs} (g)
=\la n, r\vert {\hat g} \vert n, s\ra$; it is the matrix representative of the group element
$g$ in this representation.
One can then check easily that
\beq
g^{\alpha_1}_{~k+1}g^{\alpha_2}_{~k+1}\cdots g^{\alpha_n}_{~k+1}
= \D^{(n)}_{r, w} = \la n, r\vert {\hat g} \vert n, w\ra
\label{fs23}
\eeq
where the state $\vert n, w\ra$ is the lowest weight state
obeying
\beq
\begin{split}
T_{k^2+2k} ~\vert n, w\ra &= - n {k \over \sqrt{2k (k+1)}}~ \vert n, w\ra \\
T_a ~\vert n, w\ra &= 0\\
\end{split}
\label{fs24}
\eeq
Here $T_a$ are the generators of the $SU(k)$ subalgebra and $T_{k^2 +2k}$ is the generator of the $U(1)$ algebra, both for the subgroup $U(k)$
of $SU(k+1)$.
The normalized wave functions for the basis states are thus
$\Psi_r = \sqrt{N} ~ \D^{(n)}_{r,w}(g)$.  Notice that
$\D^{(n)}_{n,w}$ are invariant under right translations of
$g$ by $SU(k)$ transformations, and under the $U(1)$ defined by
$T_{k^2+2k}$ they have a definite charge $n$, up to the $k$-dependent normalization
factor. Since they are not $U(1)$ invariant, they are really not functions
on ${\bf CP}^k$, but sections of
a line bundle on $SU(k+1)/U(k)$, the rank of the bundle being $n$.
This is exactly what we should expect for quantization of
${\bf CP}^k$ since this space is given as
$SU(k+1)/U(k)$.
\subsection{Star products for fuzzy ${\bf CP}^k$}

As for the flat case, we can construct a star product for 
functions on ${\bf CP}^k$ which captures the 
noncommutative algebra of functions \cite{garnik, KN}.
First we need to establish some notation.
Let $t_A$ denote the generators of the Lie algebra of $SU(k+1)$, realized
as $(k+1\times k+1)$-matrices. (The $T$'s given in equation
(\ref{fs24}) correspond to the generators of $U(k)\subset SU(k+1)$
in the rank $n$ symmetric representation; they are the rank $n$ representatives of
$t_{k^2+2k}$ and $t_a$. The remaining generators are of two types,
$t_{-i}$, $i= 1, 2, ..., k$, which are lowering operators and $t_{+i}$ which are raising operators.)
Left and right translation operators
on $g$ are defined by the equations
\beq
{\hat L}_A ~g = t_A ~g , \hskip .5in {\hat R}_A ~g = g~ t_A
\label{fs25}
\eeq
If $g$ is parametrized by $\vf^A$, some of which are he $\xi$'s,
then we write
\beq
g^{-1}dg = -it_A E^A_{~B} ~d\vf^B, \hskip .5in dg g^{-1} = -it_A {\tilde E}^A_{~B} ~d\vf^B
\label{fs26}
\eeq
The operators ${\hat L}_A$ and ${\hat R}_A$ are then realized as differential operators
\beq
{\hat L}_A = i ({\tilde E}^{-1})^B_{~A} {\del \over \del \vf^B},
\hskip .5in
{\hat R}_A = i ({E}^{-1})^B_{~A} {\del \over \del \vf^B}
\label{fs27}
\eeq
The state $\vert n, w\ra$, used in $\D^{(n)}_{rw}(g)
= \la n, r\vert {\hat g} \vert n, w\ra$, is the lowest weight state, which means that we have
the condition 
\beq
{\hat R}_{-i} \D^{(n)}_{r,w} =0
\label{fs27a}
\eeq
This is essentially a holomorphicity condition. 
Notice that $f(\xi )$ are holomorphic in the $\xi$'s; the $\D^{(n)}_{rw}$ have an additional
factor $(1+{\bar\xi}\cdot \xi )^{-n/2}$, which can be interpreted as due to
the nonzero connection in ${\hat R}_{-i}$, ultimately due to the
nonzero curvature of the bundle.
Equation (\ref{fs27a}) tells us that
$\D^{(n)}_{r,w} $ are sections of a rank $n$ {\it holomorphic} line bundle.

We define the symbol corresponding to a matrix $A_{ms}$
as
the function
\beqar
A(g) &=&A(\xi, {\bar\xi})= \sum_{ms}\D^{(n)}_{m, w}
(g) A_{ms} \D^{*(n)}_{s, w}(g)\nonumber\\
  &=& \la w\vert {\hat g}^T {\hat A} {\hat g}^* \vert w\ra
\label{fs28}
\eeqar
The
symbol corresponding to the product of two matrices
$A$ and $B$ can be simplified as follows.
\beqar
(AB)&=& \sum_r A_{mr} B_{rs} \D^{(n)}_{m, w}
(g) \D^{*(n)}_{s, w}(g) \nonumber\\
&=& \sum_{rr'p} \D^{(n)}_{m, w}(g)~ A_{mr}
~\D^{*(n)}_{r,p}(g) \D^{(n)}_{r',p}(g)~
B_{r's} ~\D^{*(n)}_{s, w}(g)
\label{fs29}
\eeqar
where we use the fact that $g^* g^T =1$, which reads in the rank $n$ symmetric representation
as $\delta_{rr'}=\sum_p
\D^{*(n)}_{r,p}(g)
\D^{(n)}_{r',p}(g)$.
In the sum over $p$ on the right hand side of
(\ref{fs29}), the term with $p =- n$ (corresponding to the lowest weight state
$\vert n, w\ra$)
gives the product of the symbols for $A$ and $B$.
The terms with $p > -n$ may be written in terms of powers of the raising
operators $R_{+1}$, $R_{+2}, \cdots$, $R_{+k}$, as
\beq
\D^{(n)}_{r,p}(g) = \left[{(n-s)!\over n! i_1! i_2! \cdots i_k!}
\right]^{1\over 2} {\hat R}_{+1}^{i_1}~{\hat R}_{+2}^{i_2} \cdots
{\hat R}_{+k}^{i_k} ~\D^{(n)}_{r,w}(g)\ .
\label{fs30}
\eeq
Here $s=i_1+i_2+\cdots +i_k$
and the eigenvalue for the $U(1)$ generator
$T_{k^2+2k}$ for the state
$\vert n, p\ra$ is $(-nk +sk +s )/\sqrt{2k(k+1)}$.

We also get
\beq
\biggl[{\hat R}_{+i}\D^{(n)}_{r',w}(g)\biggr] B_{r's} \D^{*(n)}_{s,  
w} (g)
= \biggl[{\hat R}_{+i}
\D^{(n)}_{r',w}
B_{r's} \D^{*(n)}_{s, w} (g)\biggr] = {\hat R}_{+i} B(g)\ .
\label{fs31}
\eeq
where we used the fact that ${\hat R}_{+i} \D^{*(n)}_{s, w} =0$.
Keeping in mind that ${\hat R}_+^* = - {\hat R}_-$,
equations (\ref{fs29} -\ref{fs31}) combine to give
\beqar
(AB)(g) &=& \sum_s (-1)^s \left[ {(n-s)! \over n! s!}\right]~
\sum_{i_1+\cdots +i_k=s}^n~
{s! \over i_1! i_2! \cdots i_k!}\nonumber\\[1mm]
&&~~~~\times{\hat R}_{-1}^{i_1} {\hat R}_{-2}^{i_2}
\cdots {\hat R}_{-k}^{i_k} A(g)~
{\hat R}_{+1}^{i_1} {\hat R}_{+2}^{i_2}\cdots {\hat R}_{+k}^{i_k}
B(g)\nonumber\\
&\equiv& A(g) * B(g)
\label{fs32}
\eeqar
This expression gives the
star product for functions on ${\bf CP}^k$. 
As expected, the first term of the sum on the right hand side gives 
the ordinary product $A(g) B(g)$,
successive terms involve derivatives and are down by powers of
$n$, as $n\rightarrow \infty$. Since the dimension of the matrices
is given by $N = (n+k)! /n! k!$, the large $n$ 
is what we need for the  limit of the continuous manifold, and the star product, as written here, is suitable for
extracting this limit for various quantities.
For example, for the symbol corresponding
to the commutator of $A, ~B$, we have
\beqar
\bigl( [A,B]\bigr)(g) &=& -{1\over n}
\sum_{i=1}^{k}({\hat R}_{-i} A ~{\hat R}_{+i} B - {\hat R}_{-i} B~ {\hat
R}_{+i} A )
~+~ {\mathcal O} (1/n^2) \nonumber\\
&=& {i\over n} \{ A, B\} ~+~ {\mathcal O}(1/n^2)\label{fs33}
\eeqar
The term involving the action of ${\hat R}$'s on the functions can indeed be verified to be
the Poisson bracket on ${\bf CP}^k$.
Equation (\ref{fs33}) is again the general correspondence of commutators and Poisson brackets,
here realized for the specific case of
${\bf CP}^k$.

We also note that traces of matrices can be converted to phase space integrals.
For a single matrix $A$, and for the product of two matrices $A, ~B$, we find
\beqar
\Tr A &=& \sum_m {A}_{mm} = N~\int d\mu (g)
\D^{(n)}_{m, w} ~A_{mm'}~
\D^{*(n)}_{m', w}\nonumber\\
&=& N \int d\mu (g)~ A(g)\nonumber\\
\Tr AB&=& N \int d\mu (g) ~A(g)*B(g)
\label{fs34}
\eeqar

\subsection{The large $n$-limit of matrices}

Consider the symbol for the product ${\hat T}_B {\hat A} $, where
${\hat T}_B$ are the generators of $SU(k+1)$,
viewed as linear operators on the states.
We can simplify it along the following lines.
\beqar
({\hat T}_B {\hat A} )_{rs} &=& \la r \vert~ {\hat g}^T~
{\hat T}_B ~{\hat A}~ {\hat g}^*
~\vert s \ra \nonumber\\
&=& S_{BC} ~\la r \vert ~{\hat T}_C ~{\hat g}^T ~{\hat A}~ {\hat  
g}^*
~\vert s \ra
\nonumber\\
&=& S_{Ba} (T_a)_{rp}~ \la p \vert ~{\hat g}^T ~{\hat A}~
{\hat g}^* ~\vert s \ra
+ S_{B+i} ~\la r \vert~ {\hat T}_{-i}~ {\hat g}^T ~{\hat A}~
{\hat g}^* ~\vert s \ra\nonumber\\[1mm]
&&\hskip .05in
+ S_{B~k^2+2k}~ \la r \vert~ {\hat T}_{k^2+2k}~ {\hat g}^T~ {\hat
A} ~{\hat g}^* ~\vert s \ra
\nonumber\\
&=& {\mathcal L}_{B}~ \la r \vert ~{\hat g}^T ~{\hat  
A}~
{\hat g}^* ~\vert s \ra
\nonumber\\
&=& {\mathcal L}_{B }~ A(g)_{rs}
\label{fs35}
\eeqar
where we have used ${\hat g}^T {\hat T}_B {\hat g}^* = S_{BC} {\hat  
T}_C$,
$S_{BC} = 2 \Tr (g^T t_B g^* t_C)$.
(Here $t_B, t_C$ and the trace are in the fundamental representation of
$SU(k+1)$.)
We have also used the fact that the states $\vert r \ra$, $\vert  
s\ra$
are $SU(k)$-invariant. (They are both equal to $\vert n, w\ra$, but we will
make this identification only after one more step of simplification.)
${\mathcal L}_B $ is defined as
\beq
{\mathcal L}_{B}=
-{nk\over \sqrt{2k(k+1)}}S_{B ~k^2 +2k}~
~ +~~ S_{B+i} {\hat{{\tilde R}}}_{-i}
\label{fs36}
\eeq
and ${\hat{{\tilde R}}}_{-i} $ is a differential operator
defined by $\hat{{\tilde R}}_{-i} g^T
= T_{-i} g^T $; it can be written in terms of  ${\hat R}_{-i}$ but the precise formula
is not needed here.

By taking ${\hat A}$ itself as a product of $\hat{T}$'s, we can iterate this calculation
and obtain the 
symbol for any product of $\hat{T}$'s as
\beq
({\hat T}_A {\hat T}_B \cdots {\hat T}_M)
= {\mathcal L}_{A} {\mathcal L}_{B  
}\cdots
{\mathcal L}_{M} \cdot 1\ .
\label{fs37}
\eeq
where we have now set $\vert n,r \ra = \vert n, s\ra = \vert n,w\ra$.

A function on fuzzy ${\bf CP}^k$ is an $N\times N$-matrix.
It can be written as a linear combination of products of ${\hat T}$'s,
and by using the above formula, we can obtain its large $n$ limit.
When $n$ becomes very large, the term that dominates in ${\mathcal L}_A$
is $S_{A~k^2+2k}$.
We then see that for any matrix function we have the relation,
$F({\hat T}_A ) \approx F(S_{A~k^2+2k})$.

We are now in position to define a set of ``coordinates'' 
$X_A$ by
\beq
X_A = -{1\over \sqrt{C_2(k+1,n)}}~{ T}_A\label{fs38}
\eeq
where $T_A$ is the matrix corresponding to ${\hat T}_A$ and 
\beq
C_2( k+1 ,n)  = {n^2k^2\over 2k (k+1)} + {nk\over 2}
\label{fs39}
\eeq
is the value of the quadratic Casimir for the
symmetric rank $n$ representation.
The coordinates  $X_A$  are $N\times
N$-matrices and can be taken as
the coordinates of fuzzy ${\bf CP}^k$, embedded
in ${\bf R}^{k^2+2k}$. In the large $n$ limit, we evidently have
$X_A \approx S_{A ~k^2+2k} = 2 \Tr (g^T t_A g^* t_{k^2+2k})$.
From the definition, we can see that $S_{A ~k^2+2k}$ obey algebraic constraints
which can be verified to be the correct ones for
describing ${\bf CP}^k$ as embedded in
${\bf R}^{k^2+2k}$ \cite{NR1}.

\section{Noncommutative plane, fuzzy ${\bf CP}^1$, ${\bf CP}^2$, etc.}
\label{sec:2}

The noncommutative plane has already been described. The basic
commutation rules are given by (\ref{fs2}), with the indices taking
only one value, $1$. The star product is given by (\ref{fs10a}).
While the coherent state basis is very ideal for considering the
commutative limit $\theta \rightarrow 0$, for many purposes, it is easy enough
to deal with the representation of $Z, \bZ$ as infinite-dimensional 
matrices. In fact, one can also use real coordinates and characterize them by 
the commutation rules
\beq
[X_i , X_j ] = i \theta ~\epsilon_{ij}
\label{fs39a}
\eeq
More generally, one may consider ${\bf R}^{2k}$, with the commutation rules
\beq
[ X_i, X_j ] = i~ \theta_{ij}
\label{fs39b}
\eeq
where the constant matrix $\theta_{ij}$ characterizes the noncommutativity.

Fuzzy ${\bf CP}^1$ is the same as the fuzzy two-sphere and has been studied for a
long time \cite{madore}.
It can be treated as the special case $k=1$ of our analysis.
The Hilbert space corresponds to representations of 
$SU(2)$, and they are
given by the standard angular momentum theory.
Representations are labeled by the maximal angular momentum $j =  
{n\over 2}$,
with $N =2j+1 =n+1$.
The generators are the angular momentum matrices, and the
coordinates of fuzzy $S^2$ are given by $X_i = J_i /\sqrt{j(j+1)}$, 
as in equation(\ref{fs38}).
These coordinate matrices obey the commutation rule
\beq
[X_i , X_j ] = {i\over \sqrt{j(j+1)}}\, \epsilon_{ijk} X_k\ . \label{fs40}
\eeq
We get commuting coordinates only at large $n$. 

If $g$ is an element of $SU(2)$ considered as a $2\times 2$-matrix, 
we can parametrize it, apart from an overall $U(1)$ factor and
along the lines of (\ref{fs21}), as
\beq
g = {1\over \sqrt{(1+{\bar\xi}\xi )}} \left[ \begin{matrix}-1&~\xi \\
~{\bar \xi}&~1\\
\end{matrix}\right]\,.
\label{fs41}
\eeq
The large $n$ limit of the coordinates is given by
$X_i \approx S_{i3}(g)$, which can be worked out as
\beqar
S_{13}= - {\xi +{\bar\xi} \over (1+\xi{\bar\xi} )}\ , &\hskip .3in& S_{23} = -i \,{{\xi  
-{\bar\xi}}
\over (1+\xi{\bar\xi})}\ ,
\hskip .3in
S_{33} = {\xi{\bar\xi} -1 \over \xi{\bar\xi} +1}\ .
\label{fs42}
\eeqar
The quantities $S_{i3}$ obey the condition $S_{i3}S_{i3} =1$
corresponding to a unit two-sphere embedded in ${\bf R}^3$;
$\xi, ~{\bar\xi}$ are the local complex coordinates for the sphere.
The  matrix coordinates obey the condition $X_i X_i =1$.
Thus we may regard them as giving the fuzzy two-sphere, which approximates 
to the continuous two-sphere as $n \rightarrow \infty$.

We can also study functions on fuzzy $S^2$, which are given as $N\times N$-matrices.
At the matrix level, there are $N^2 = (n+1)^2$ independent  
``functions''. A basis for them is given by
${\bf 1}$, $X_i$, $X_{(i}X_{j)}$, etc., where $X_{(i}X_{j)}$ denotes the symmetric
part of the product $X_i X_j$ with all contractions of indices $i, j$ removed;
i.e., $X_{(i}X_{j)} = {1\over 2}(X_i X_j +X_j X_i) - {1\over 3} \delta_{ij} {\bf 1}$.
Since we have finite-dimensional matrices, the last independent function
corresponds to the symmetric $n$-fold product of $X_i$'s with all contractions
removed.

On the smooth $S^2$, a basis for functions is given by the spherical  
harmonics,
labeled by the integer $l = 0, 1, 2,...$. They are given by
the products of $S_{i3}$ with all contractions of indices removed.
There are $(2l+1)$ such functions for each value of $l$.
If we consider a truncated set of functions with a maximal value
of $l$ equal to $n$, the number of functions is
$\sum_0^n (2l+1) = (n+1)^2$. Notice that this number coincides with
the number of ``functions'' at the  matrix level.
There is one-to-one correspondence with the spherical
harmonics, for $l=0, 1, 2$, etc., up to $l=n$.
Further, by using the relation $X_i \approx S_{i3}$, we can see that the  
matrix functions, ${\bf 1}$, $X_i$, $X_{(i}X_{j)}$, etc., 
in the large $n$ limit, approximate to the
the spherical harmonics.
The set of functions at the matrix level go over  
to the
set of functions on the smooth $S^2$ as $n \rightarrow \infty$.
Fuzzy $S^2$ may thus be viewed as a regularized version of the smooth
$S^2$ where we impose a cut-off on the number of modes of a function;
$n$ is the regulator or cut-off parameter.

Fuzzy ${\bf CP}^2$ is the case $k=2$ of our general anlysis.
The coordinates are given by
\beq
X_A = - {3\over \sqrt{n(n+3)}} T_A
\label{fs43}
\eeq
The large  
$n$
limit of the
coordinates $X_A$ are $S_{A8} = 2 \Tr (g^T t_A g^* t_8)$.
In this limit, the coordinates obey the condition
\beq
\begin{split}
X_A X_A =& 1\\
d_{ABC} X_B X_C =& -{1\over \sqrt{3}}\, X_C
\end{split}
\label{fs44}
\eeq
where $d_{ABC} = 2\Tr ~t_A (t_B t_C + t_C t_B)$.
These conditions are known to be the equations for ${\bf  
CP}^2$ as embedded in ${\bf R}^8$.
Thus, our definition of fuzzy ${\bf CP}^2$ does approximate to the
smooth ${\bf CP}^2$ in the large $n$ limit. 
Equations (\ref{fs44}) can also be imposed at the level of  
matrices
to get a purely matrix-level definition of fuzzy ${\bf CP}^2$
\cite{garnik, NR1}.

The dimension of the Hilbert space is given by
$N= {1\over 2} (n+1) (n+2)$. Matrix functions are
$N\times N$-matrices; a basis for them is given by
products of the $T$'s with up to $N-1$ factors.
There are $N^2$ independent functions possible. 
On the smooth ${\bf  CP}^2$,
a basis of functions is given by 
products of the form $\bu_{\beta_1}\bu_{\beta_2}\cdots\bu_{\beta_l}
u^{\alpha_1}u^{\alpha_2}\cdots u^{\alpha_l}$, where
$u^\alpha = g^\alpha_{~3}$. 
The number of such functions, for a given value of $l$, is
\beq
\left[ {1\over 2} (l+1) (l+2)\right]^2 - \left[
{1\over 2} l (l+1)\right]^2 = (l+1)^3
\label{fs45}
\eeq
(All traces for these functions correspond to lower ones and 
can be removed from the counting at the level $l$.)
If we consider a truncated set of functions, with values of $l$ going up to
$n$, the number of independent functions will be
\beq
\sum_0^n (l+1)^3 = {1\over 4} (n+1)^2 (n+2)^2 = N^2\, .
\label{fs46}
\eeq
It is thus possible to consider the fuzzy ${\bf CP}^2$ as a  
regularization
of the smooth ${\bf CP}^2$ with a cut-off on the number of modes of a  
function.
Since any matrix function can be written as a sum of products of ${\hat  
T}$'s,
the corresponding  large $n$ limit has a sum of
products of $S_{A8}$'s.
The independent basis functions are thus given by representations of
$SU(3)$ obtained from reducing symmetric products of the adjoint  
representation
with itself. These are exactly what we expect based on the fact the smooth
${\bf CP}^2$ is given by the embedding conditions
(\ref{fs44}). The algebra of matrix functions for the fuzzy ${\bf CP}^2$,
as we have constructed it, does go over to the algebra of functions
on the smooth ${\bf CP}^2$.

Since the fuzzy spaces, the fuzzy ${\bf CP}^k$ in particular,
can be regarded as a regularization
of the smooth ${\bf CP}^k$ with a cut-off on the number of modes of a  
function,
they can be used for regularization of field theories,
in much the same way that lattice regularization of field theories is  
carried out.
There are some interesting features or fuzzy regularization; for  
example, it may be possible to evade
fermion doubling problem on the lattice \cite{fuzzfield}.

\section{Fields on fuzzy spaces, Schr\"odinger equation}
\label{sec:3}

A scalar field on a fuzzy space can be written as $\Phi (X)$, indicating  
that it is
a function of the coordinate matrices $X_A$. Thus $\Phi$ is an
$N\times N$-matrix.
Further, equation (\ref{fs33}) tells us that
\beqar
[T_A , \Phi ] &\approx& - {i \over n}\, {nk \over \sqrt{2k (k+1)}}\, \{
S_{A~k^2+2k},
\Phi \}\nonumber\\[1mm]
&\equiv&-i D_A \Phi \ .\label{fs47}
\eeqar
$D_A$, as defined by this equation, are the derivative operators
on the space of interest. For example, 
for the fuzzy $S^2$, they are given by
\beq
\begin{split}
D_1 &= {1\over 2}\, ( {\bar\xi}^2 \del_{{\bar\xi}} +\del_\xi - \xi^2 \del_\xi
-\del_{{\bar\xi}})\\
D_2 &= -{i\over 2}\, ( {\bar\xi}^2 \del_{{\bar\xi}} + \del_\xi +\xi^2 \del_\xi
+\del_{{\bar\xi}})\\
D_3 &= {\bar\xi} \del_{{\bar\xi}} - \xi \del_\xi
\end{split}
\label{fs48}
\eeq
These obey the $SU(2)$ algebra, $[D_A ,D_B] = i \epsilon_{ABC} D_C$.
They are the translation operators on the two-sphere and correspond to
the three
isometry transformations. 

These equations show that we can define the derivative, at the matrix level, as
the commutator $i [T_A, \Phi ]$,
which is the adjoint action of $T_A$ on $\Phi$.
The Laplacian on $\Phi$ is then given by
$- \Delta \cdot \Phi = [T_A, [T_A, \Phi ]]$.
The Euclidean action for a scalar field can be taken as
\beq
{\mathcal S} = {1\over N} \Tr \biggl[  \Phi^\dagger  [T_A, [T_A, \Phi  
]]  +
V (\Phi )\biggr]
\label{fs49}
\eeq
where $V(\Phi )$ is a potential term; it does not involve derivatives.

The identification of derivatives also lead naturally to
gauge fields.
We introduce a gauge field ${\mathcal A}_A$ by defining the covariant derivative
as
\beq
-i \D_A \Phi = [T_A , \Phi ] + {\mathcal A}_A \Phi
\label{fs50}
\eeq
where ${\mathcal A}_A$ is a set of hermitian matrices. In the absence  
of the
gauge field, we have the commutation rules $[T_A, T_B] =i f_{ABC}  
T_C$. The 
field strength tensor ${\mathcal F}_{AB}$, which is the deviation from this
algebra, is thus given by
\beq
-i {\mathcal F}_{AB} =  [ T_A +{\mathcal A}_A , T_B +{\mathcal A}_B] -
if_{ABC} (T_C +{\mathcal A}_C)\ .
\label{fs51}
\eeq
The action for Yang-Mills theory on a fuzzy space is then given by
\beq
{\mathcal S} = {1\over N}\, \Tr \biggl[ {1\over 4}\,{\mathcal F}_{AB}  
{\mathcal
F}_{AB}
\biggr] \label{fs52}
\eeq

The quantum theory of these fields can be defined by the functional integral
over actions such as (\ref{fs49}) and (\ref{fs52}).
Perturbation theory, Feynman diagrams, etc., can be worked out.
Our main focus will be on  particle dynamics, so we will not do this here.
However, some of the relevant literature can be traced from
\cite{douglas, fuzzfield, others2}.

One can also write down the Schr\"odinger equation for particle 
quantum mechanics on a fuzzy space \cite{PN}.
The wave function $\Psi (X)$ is matrix and its derivative is given by
$-iD_A \Psi =  [T_A , \Psi] $.
Coupling to an external potential may be taken to be
of the form $V(X) \Psi$. The 
Schr\"odinger equation is then given by
\beq
i {\del \Psi \over \del t} ~+~ D_A ( D_A \Psi )~-~ V \Psi
=0
\label{fs53}
\eeq
When it comes to gauge fields, there is a slight subtlety. 
The covariant derivative is of the form
(\ref{fs50}). To distinguish the action
of the gauge field from the potential $V$, for the 
covariant derivative we may use the definition
$-i \D_A \Psi = [T_A , \Psi ] + \Psi {\mathcal A}_A $.
(One could also change the action of the potential.)
The Schr\'odinger equation retains the usual form,
\beq i\D_0 \Psi + {1\over 2m}~\D_A (\D_A \Psi ) - V \Psi =0
\label{fs54}
\eeq 

\section{The Landau problem on ${\bf R}^2_{NC}$ and ${\bf S}^2_F$}
\label{sec:4}
As a simple example of the application of the ideas given above, we shall now work out the
quantum mechanics of a charged particle in a magnetic field on the fuzzy two-plane
\cite{PN, duval}. This is the fuzzy version of the classic Landau problem. We shall also include an oscillator potential
to include the case of an ordinary potential as well.
At the operator level, the inclusion of a background magnetic field is easily achieved by changing
the commutation rules for the momenta.
The modified algebra of observables is given by
\beq
\begin{split}
[ X_1, X_2] &= i~ \theta\\
[X_i , P_j] &= i~ \delta_{ij}\\
[ P_1, P_2]&= i ~B
\end{split}
\label{lan1}
\eeq
where $i, j = 1, 2$, and $B$ is the magnetic field.
The Hamiltonian may be taken as
\beq 
H= {1\over 2}\left[ P_1^2 +P_2^2 +\omega^2 (X_1^2 +X_2^2
)\right]
\label{lan2}
\eeq 
We have chosen the isotropic oscillator (with frequency $\omega$), 
and $H$ is invariant under rotations.
The form of various operators can be slightly different from the usual
ones because of the noncommutativity
of the coordinates. The angular momentum is given by
\beq 
L=\frac{1}{1-\theta B} \left[ X_1 P_2 - X_2 P_1 +
\frac{B}{2} (X_1^2 + X_2^2 ) +
\frac{\theta}{2} (P_1^2 + P_2^2 )\right]
\label{lan3}
\eeq
$L$ commutes with $H$, as can be checked easily.

The strategy for solving this problem involves expressing $X_i, P_i$ in terms of a
a usual canonical set, so that thereafter, it can be treated as an ordinary quantum mechanical system. This change of variables will be different
for $B < 1/\theta$ and for $B>1/\theta$. For $B<1/\theta$, we define a change of
variables
\beq
\begin{split}
 X_1 =l \alpha_1, \hskip .3in&\hskip .3in P_1= {1\over
l}\beta_1 +q \alpha_2\\
X_2= l \beta_1, \hskip .3in &\hskip .3in 
P_2= {1\over l}\alpha_1 -q \beta_2
\end{split}
\label{lan4}
\eeq
where $l^2=\theta$ and $q^2 = (1- B\theta )/\theta$.
$\alpha_i, \beta_i$ form a standard set of canonical variables, with
\beq
\begin{split}
[ \alpha_i , \alpha_j ]&=0\\
[\alpha_i ,\beta_j] &=i~\delta_{ij}\\
[\beta_i, \beta_j]&=0
\end{split}
\label{lan5}
\eeq
The Hamiltonian is now given by
\beq
 H= {1\over 2}\left[
\left(\omega^2 l^2 +{1\over l^2}\right) (\alpha_1^2
+\beta_1^2) +q^2(\alpha_2^2 +\beta_2^2) +{2q\over l}
(\alpha_1 \beta_2 + \alpha_2 \beta_1)\right]
\label{lan6}
\eeq
We can eliminate the mixing of the two sets of variables in the last term, and diagonalize $H$, by making a Bogoliubov transformation which will express
$\alpha_i, \beta_i$ in  terms of  a canonical set
$q_i,~p_i$ as
\beq
\left(\begin{matrix}\alpha_1\\
\alpha_2\\
\beta_1\\
\beta_2\\ \end{matrix}\right)= \cosh \lambda \left(\begin{matrix} q_1\\ q_2\\
p_1\\ p_2\\ \end{matrix} \right) ~+~ \sinh\lambda \left(\begin{matrix} p_2\\
p_1\\ q_2\\ q_1\\ \end{matrix}
\right)\label{lan7}
\eeq 
The Hamiltonian can be diagonalized by the choice
\beq
\tanh 2\lambda = -~{2ql \over 1+ \omega^2 l^4 + q^2
l^2}\label{lan8}
\eeq 
and is given by
\beq H= {1\over 2} \left[ \Omega_+ ~(p_1^2+q_1^2)+\Omega_{-}
(p_2^2 +q_2^2)\right]
\label{lan9}
\eeq
with
\beq
\Omega_{\pm} = {1\over 2} \sqrt{(\omega^2\theta-B)^2 +4\omega^2}
~\pm {1\over 2} (\omega^2 \theta +B) \label{lan10}
\eeq 
From equation (\ref{lan9}), we see that the problem is
equivalent to that of two harmonic oscillators with
frequencies
$\Omega_+$
and $\Omega_{-}$.

The case of $B>1/\theta$ can be treated in a similar way. We again have two oscillators,
with the Hamiltonian (\ref{lan9}), but with frequencies given as
\beq
\Omega_{\pm} = \pm {1\over 2} \sqrt{(\omega^2\theta-B)^2
+4\omega^2} ~+ {1\over 2} (\omega^2 \theta +B) \label{15}
\eeq 
Notice that $B\theta =1$ is a special value, for both regions of $B$,
with one of the frequencies 
becoming zero. The symplectic two-form which leads to the commutation rules
(\ref{lan1}) is given by
\beq
\Omega = {1\over 1-B\theta } \left(
dP_1 dX_1 + dP_2 dX_2 +  \theta dP_1 dP_2 + B dX_1 dX_2
\right)
\label{lan12}
\eeq
The phase space volume is given by
\beq
d \mu = {1\over \vert 1-B\theta\vert}~d^2X d^2P
\label{lan13}
\eeq
A semiclassical estimate of the number of states is given by the volume divided by
$(2\pi )^2$. The formula (\ref{lan13}) shows  that the density of states
diverges at $B\theta =1$, again indicating that it is a special value.

The Landau problem on the fuzzy sphere can be formulated in a similar way.
On the sphere, the translation operators are the angular momenta $J_i$
and the algebra of observables is given by
\beq
\begin{split}
[X_i , X_j]  &=0\\
[J_i, X_j]&= i \ep_{ijk}X_k\\
[J_i, ~J_j ]&= i \ep_{ijk} J_k
\end{split}
\label{lan14}
\eeq
$X^2$ commutes with all operators and its value can be fixed to be
$a^2$, where $a$ is the radius of the sphere. The other Casimir operator
is $X\cdot J $; its value is written as $- a (n/2)$, where $n$ must be an integer and gives
the strength of the magnetic field; it is the charge of the
monopole at the center of the sphere 
(if we think of it as being embedded in ${\bf R}^3$).

For the fuzzy case, the coordinates themselves are noncommuting and 
are given, up to normalization, by $SU(2)$ operators $R_i$ as
$ X_i = a R_i /\sqrt{C_2}$.
The algebra of observables becomes
\beq
\begin{split}
[R_i , R_j]  &=i\ep_{ijk}R_k\\
[J_i, R_j]&= i \ep_{ijk}R_k\\
[J_i, ~J_j ]&= i \ep_{ijk} J_k
\end{split}
\label{lan15}
\eeq
The Casimir operators are now $R^2$ and $R\cdot J - {1\over 2}J^2$, the latter
being related to the strength of the magnetic field.
The algebra (\ref{lan15}) can be realized by two independent $SU(2)$ algebras
$\{ R_i \}$ and $\{ K_i \}$, 
with $J_i = R_i +K_i$. The two Casimirs are now $R^2$ and $K^2$, which we fix to the values
$r(r+1)$ and $k(k+1)$, $r$, $k$ being positive half-integers.
The difference $k -r= n/2$. The limit of the smooth sphere
is thus obtained by taking
$k, r \rightarrow \infty$, with $k-r$ fixed. As the generalization of $P^2/2m$, we take the Hamiltonian as
\beq
H = {\gamma \over 2 a^2}~J^2
\label{lan16}
\eeq
where $\gamma$ is some constant. The spectrum of the Hamilonian is now easily calculated
as
\beq
E = {\gamma \over 2a^2} ~ j (j+1), \hskip .5in j = {\vert n \vert \over 2},
 {\vert n \vert \over 2}+1, \cdots , j+k
 \label{lan17}
 \eeq
From the commutation rule for the coordinates $X_i = a R_i /\sqrt{r (r+1)}$, we may identify the noncommutativity parameter
as $\theta \approx a^2/r$, for large $r$.
The limit of this problem to the noncommutative plane can be obtained
by taking $r$ large, but keeping $\theta$ fixed. Naturally, this will require a large radius
for the sphere.
The strength of the magnetic field in the plane is related to $n$ by
$ (1- B\theta )n = 2Ba^2$. For more details, see \cite{PN}; 
also the Landau problem on general noncommutative Riemann surfaces has been 
analyzed, see \cite{MP}.

\section{Lowest Landau level and fuzzy spaces}
\label{sec:5}

There is an interesting connection between the Landau problem on
a smooth manifold $M$ and the construction of
the fuzzy version of $M$; we shall explain this now.

The splitting of Landau levels is controlled by the magnetic field
and, if the field is sufficiently strong, transitions between levels 
are suppressed
and the dynamics is restricted to one level, say, the lowest.
The observables are given as
hermitian operators on this subspace of the Hilbert space 
corresponding to the lowest Landau level;
they can be
obtained by projecting the full operators to this subspace. 
The commutation rules can change due to this projection.
The position coordinates, for example, when projected to
the lowest Landau level (or any other level), are no longer
mutually commuting. 
The dynamics restricted to the lowest Landau
level is thus dynamics on a noncommutative space.
In fact, the Hilbert (sub)space of the lowest Landau level can be
taken as the Hilbert space ${\cal H}_N$ used to define the fuzzy version
of $M$. Thus the solution of the Landau problem on smooth $M$
gives a construction of the fuzzy version of $M$.

We can see how this is realized explicitly by analyzing the two-sphere
\cite{haldane}.
Since $S^2 = SU(2)/U(1)$, the wave functions can be obtained in terms of
 functions on the group
$SU(2)$, i.e., in terms of the Wigner functions $\D^{(j)}_{rs}(g)$.
We need two derivative operators which can be taken as 
two of the right translations of $g$, say, 
$R_{\pm}=R_1 \pm i R_2$. With the correct dimensions, the covariant derivatives can be written
as
\beq 
D_{\pm} =
i~{R_{\pm} \over a}\label{qh1}
\eeq
The $SU(2)$ commutation rule $ \bigl[ R_{+},~R_{-}
\bigr] = 2~ R_{3}$ shows that the covariant derivatives do not commute and we
may identify the value of $R_3$ as the field strength.
In fact, comparing this commutation rule to $[D_+ ,D_-] = 2B$,  
we see that $R_{3}$ should be taken to be $- (n/2)$, where $n$ 
is the monopole number, $n =2Ba^2$. Thus the wave functions
on $S^2$ with the magnetic field background are of the form
$\Psi_m \sim {\D}^{(j)}_{m,-{n \over 2}} (g)$. 

The one-particle Hamiltonian is given by
\beq
H = - {1 \over {4\mu}} \bigl( D_{+} D_{-} + D_{-} D_{+} \bigr)  
= {1 \over {2\mu a^2}}\, \Bigl( \sum_{A=1}^3 R_{A}^2 - R_3^2 \Bigr)
\label{qh2}
\eeq
where $\mu$ is the particle mass.
The eigenvalue $-{n\over 2}$ must occur as one of the possible values for
$R_3$, so that we can form ${\D}^{(j)}_{m,-{n \over 2}} (g)$.
This means that $j$ should be of the form
$j={\vert n\vert \over 2} + q$, $q=0,1,..$.
Since $R^2=j(j+1)$,
the energy eigenvalues are easily obtained as
\beq
E_q  =  {1 \over {2\mu a^2 }} \left[ \left( {n\over 2}+ q\right) \left({n\over 2} + q +1\right) -  
{n^2
\over 4} \right] 
 =  {B \over {2\mu}}\, (2 q +1) + {{q(q+1)} \over {2\mu a^2}}\  
\label{qh3}
\eeq
The integer $q$ is the Landau level index, $q=0$ being the lowest
energy state or the ground state. The gap between levels increases as
$B$ increases, and,
in the limit of large magnetic fields, 
it is meaningful to restrict dynamics to one
level, say the lowest, if the available excitation energies are
small compared to $B/2\mu$. 
In this case, $j ={\vert n\vert \over 2}$, $R_3
=-{n\over 2}$, so that we have the lowest weight state for the right
action of $SU(2)$, taking $n$ to be positive.
The condition for the lowest Landau level is
$R_- \Psi =0$.

The Hilbert space of the
lowest Landau level 
is spanned by $ \Psi_m\sim \D^{({n\over
2})}_{m,-{n\over 2}}$.
Notice that this is exactly the Hilbert space for fuzzy $S^2$.
Hence all observables for
the lowest Landau level correspond to the observables of the fuzzy
$S^2$. 

This correspondence can be extended to the Landau problem
on other
spaces, say, ${\bf CP}^k$ with a $U(1)$ background field,
for example.
The background field specifies the
choice of the representation of $T_a$ and $T_{k^+2k}$, the $U(k)$ subalgebra
of $SU(k+1)$, 
in the Wigner ${\mathcal D}$-functions.
For zero $SU(k)$ background field, $T_a \Psi =0$ and the eigenvalue
of $T_{k^2+2k}$ gives the magnetic field, which must obey appropriate
quantization conditions. In fact, we get the equations (\ref{fs24}) and 
the Hilbert subspace
of the lowest Landau level is the same as the Hilbert space
$\H_N$ used for the construction of fuzzy ${\bf CP}^k$ \cite{KNR, KN}. 

The lowest Landau level wave functions are holomorphic, except possibly for
a common prefactor, which has to do with the inner product.
This is also seen from (\ref{fs19}). In fact, the condition (\ref{fs27a}), namely,
$R_{-i}\Psi =0$, which selects the lowest level, are the holomorphicity conditions.
The higher levels are not necessarily holomorphic.
This will be useful later in writing the Yang-Mills amplitudes in terms of a
Landau problem on ${\bf CP}^1 = S^2$.

\section{Twistors, supertwistors}
\label{sec:6}

\subsection{The basic idea of twistors}

The idea of twistors is due to Roger Penrose, many years ago,  in 1967 \cite{penrose}.
There are many related ways of thinking about twistors, but a simple approach is in terms of constructing solutions to massless wave equations or their Euclidean counterparts.

We start by considering the two-dimensional Laplace equation, which may be written in complex coordinates as
\beq
\del~ {\bdel }~f =0
\label{tw1}
\eeq
where $z= x_1 +i x_2$. The solution is then obvious, $f (x) = h(z) ~+~ g (\bz )$,
where $h(z)$ is a holomorphic function of $z$ and $g(\bz )$ is antiholomorphic. For a given physical problem such as electrostatics or two-dimensional hydrodynamics, we then have to simply guess the holomorphic function with the required singularity structure.
Further, the problem has conformal invariance and one can use the techniques of conforml mapping to simplify the problem.

We now ask the question: Can we do an analogous trick to find solutions of the four-dimensional problem, say, the Dirac or Laplace equations on $S^4$? 
Clearly this is not so simple as in two dimensions,  there are some complications.
First of all, $S^4$ does not admit a complex structure. Even if we consider ${\bf R}^4$, which is topologically equivalent to $S^4$ with a point removed, there is no natural choice of complex coordinates.
We can, for example, combine the four coordinates into two complex ones as in
\beq
\left( \begin{matrix}z_1\\
 z_2\\
 \end{matrix}\right) = \left( \begin{matrix}x_1 +i x_2\\
 x_3 +i x_4\\
 \end{matrix}\right)
\label{tw2}
\eeq
Equally well we could have considered
\beq
\left( \begin{matrix}z'_1\\
 z'_2\\
 \end{matrix}\right) = \left( \begin{matrix} x_1 +i x_3\\
  x_2 +i x_4\\
\end{matrix}\right)
\label{tw3}
\eeq
or, in fact, an infinity of other choices. Notice that any particular choice will destroy the
overall $O(4)$-symmetry of the problem. We may now ask: How many inequivalent choices can be made, subject to, say, preserving
$x^2 = {\bz}_1 z_1 + {\bz}_2 z_2$? Given one choice, as in (\ref{tw1}), we can do an $O(4)$ rotation of $x_\mu$ which will generate other possible complex combinations with the same value of $x^2$.  However, if we do a $U(2)$-transformation of $(z_1, z_2)$, this gives us a new combination of the $z$'s preserving holomorphicity. In particular, a holomorphic function of the $z_i$ will remain a holomorphic function after a $U(2)$ rotation. Thus the number of inequivalent choices of local complex structure is given by $O(4)/ U(2) = S^2 = {\bf CP}^1$.
The idea now is to consider $S^4$ with the set of all possible local complex structures at each point, in other words, a ${\bf CP}^1$ bundle over $S^4$. This bundle is ${\bf CP}^3$.
The case of ${\bf R}^4$ is similar to considering a neighborhood of $S^4$.

An explicit realization of this is as follows. We represent ${\bf CP}^1$ by a two-spinor $U^A$,
$A =1,2$, with the identification $u^a \sim \lambda U^A$, where $\lambda \in {\bf C} - \{ 0 \}$.
We now take a $4$-spinor with complex element $Z^\alpha$, $\alpha = 1, 2, 3, 4$, and write it  as $Z^\alpha = (W_{\dot A}, U^A)$, where $U^A$ describes ${\bf CP}^1$ as above. The relation between $W_{\dot A}$ and $U^A$ is taken as
\beq
W_{\dot A} = x_{{\dot A} A}~U^A \label{tw4}
\eeq
$x_{{\dot A}A}$ defined by this equation may be taken as the local coordinates on $S^4$.
We can even write this out as
\beq
x_{{\dot A}A} = \left( \begin{matrix}~x_4+i x_3 &~ x_2 +i x_1\\
-x_2 +ix_1 &~ x_4-ix_3\\
\end{matrix}\right) = x_\mu e^\mu
\label{tw5}
\eeq
with $e^i = \sigma^i$ are the Pauli matrices and $e^4= {\bf 1}$, so that $x_\mu$ are the usual coordinates. One can read the equation (\ref{tw4}) in another way, namely, 
as combining the $x$'s into complex combinations $W_1 $ and $W_2$, in a manner specified by the choice of $U^A$. Thus a point on ${\bf CP}^1$, namely, a choice of $U^A$, gives a specific combination of complex coordinates. 
 
We have the identification $Z^\alpha \sim \lambda Z^\alpha$, which follows from
$U^A \sim \lambda U^A$ and the definition of $x_{{\dot A}A}$ as in (\ref{tw4}).
This means that $Z^\alpha$ define ${\bf CP}^3$. Further the indices ${\dot A},~A$ correspond to $SU(2)$ spinor indices, right and left, in the splitting $O(4) \sim SU_L(2)\times SU_R(2)$.
$Z^\alpha$ are called twistors.

Given the above-described structure, there is a way of constructing solutions to massless wave equations (or their Euclidean versions), in terms of holomorphic functions defined
on a neighborhood of ${\bf CP}^3$.
Evidently, preserving the $O(4)$ symmetry requires some sort of integration over all
$u$'s, consistent with holomorphicity. There is a unique holomorphic differential
we can make out of the $u$'s which is $O(4)$ invariant, namely, 
$U\cdot dU = \epsilon_{AB} U^A dU^B$.
We will now do a contour integration of holomorphic functions using this. Let $f(Z)$ be aholomorphic function of $Z^\alpha$ defined on some region in twistor space.
We can then construct the contour integral
\beq
{\tilde f}^{A_1 A_2 \cdots A_n} (x) =
\oint_C  U\cdot dU~ U^{A_1} U^{A_2} \cdots U^{A_n} ~f(Z) \label{tw6}
\eeq
For this to make sense on a neighborhood of ${\bf CP}^3$, $f(Z)$ should have
degree of homogeneity $-n -2$, so that the integrand is invariant under the scaling
$Z^\alpha \rightarrow \lambda Z^\alpha$, $U^A \rightarrow \lambda U^A$, and thus projects down to a proper differential on ${\bf CP}^3$. The contour $C$ will be taken to enclose some of the poles of the function $f(Z)$.  Since we write $W_{\dot A} = x_{{\dot A}A} U^A$,
after integration, we are left with a function of the $x$'s; ${\tilde f}$ is a function of
the $S^4$ or ${\bf R}^4$ coordinates; it is also a multispinor of $SU_L(2)$.

Consider now the action of the chiral Dirac operator on this, namely,
$\epsilon_{CA_1}\nabla^{{\dot B}  C} {\tilde f}^{A_1 A_2 \cdots A_n}$.
 Since $x_\mu$ appear in $f(Z)$ only via the combination
 $x_{{\dot A}A} U^A$, we can write
 \beqar
 \epsilon_{CA_1}\nabla^{{\dot B}  C} {\tilde f}^{A_1 A_2 \cdots A_n}
 &=&\epsilon_{CA_1} \oint_C U\cdot dU~ U^{A_1} U^{A_2} \cdots U^{A_n} ~~\nabla^{{\dot B}C} f(Z) \nonumber\\
 &=&\epsilon_{CA_1} \oint_C U\cdot dU~ U^{A_1} U^{A_2} \cdots U^{A_n} ~
 U^C {\del  f(Z) \over \del W_{\dot B}} \nonumber\\
 &=& 0
 \label{tw7}
 \eeqar
since $\epsilon_{CA_1} U^C U^{A_1} =0$ by antisymmetry.
Thus ${\tilde f}^{A_1 A_2 \cdots A_n}(x)$ is a solution to the chiral Dirac equation
in four dimensions.

In a similar way, one can define 
\beq
{\tilde g}^{{\dot A}_1 {\dot A}_2 \cdots {\dot A}_n}(x)
= \oint_C U\cdot dU~  {\del  \over \del W_{{\dot A}_1}}{\del  \over \del W_{{\dot A}_1}}
\cdots {\del  \over \del W_{{\dot A}_1}} ~g(Z)
\label{tw8}
\eeq
where $g (Z)$ has degree of homogeneity equal to $n-2$. It is then easy to check that
\beq
\epsilon_{{\dot B} {\dot A}_1} \nabla^{B{\dot B}}
{\tilde g}^{{\dot A}_1 {\dot A}_2 \cdots {\dot A}_n} =0
\label{tw9}
\eeq

The two sets of functions, ${\tilde f}^{A_1 A_2 \cdots A_n}(x)$ and 
${\tilde g}^{{\dot A}_1 {\dot A}_2 \cdots {\dot A}_n}(x)$, give a complete set of solutions to the chiral Dirac equation in four dimensions.
This is essentially Penrose's theorem, for this case. (The theorem is more general, applicable to other manifolds which admit twistor constructions.) The mapping between holomorphic functions in twistor space and massless fields in spacetime is known as the Penrose transform. (Strictly speaking, we are not concerned with holomorphic functions.
They are holomorphic in some neighborhood in twistor space, and further, they are not really defined on ${\bf CP}^3$, since they have nontrivial degree of homogeneity.
The proper mathematical characterization would be as sections of holomorphic sheaves of appropriate degree of homogeneity.)

\subsection{An explicit example}

As an explicit example of the Penrose transform, consider the holomorphic function
\beq
f(Z) = {1\over a\cdot W~ b\cdot W~c\cdot U}
\label{tw9a}
\eeq
where $a\cdot W = a^{\dot A} x_{{\dot A} A}U^A \equiv U^1 w^2 - U^2 w^1$, 
$b\cdot W = b^{\dot A} x_{{\dot A} A}U^A \equiv U^1 v^2 - U^2 v^1$.
Defining $z = U^2 /U^1$, we find, for the Penrose integral, 
\beqar
\psi^A &=& \oint U\cdot dU {u^A \over a\cdot W~ b\cdot W~c\cdot U}\nonumber\\
&=& \oint dz~ {U^A \over U^1} {1\over (w^2 - zw^1) (v^2 -zv^1) c^2 - zc^1)}\label{tw9b}
\eeqar
Taking the contour to enclose the pole at $w^2/w^1$, we find
\beqar
\psi^A &=& \epsilon^{AB}{ a^{\dot A} x_{{\dot A}B} \over x^2 w\cdot c} {1\over a\cdot b}
\nonumber\\
&=&  \epsilon^{AB}{ a^{\dot A} x_{{\dot A}B} \over x^2 (a x c)} {1\over a\cdot b}
\label{tw9c}
\eeqar
where $ axc = a^{\dot A} x_{{\dot A}A} c^A$. (We take $a\cdot b \neq 0$.)  One can check directly that this obeys
the equation
\beq
\nabla_{{\dot A} A} \psi^A =0
\label{tw9d}
\eeq
\subsection{Conformal transformations}

There is a natural action of conformal transformations on twistors. We can consider
$Z^\alpha$ as a four-spinor of $SU(4)$, the latter acting as  linear transformations
on $Z^\alpha$, explicitly given by
\beq
Z^\alpha ~\longrightarrow~ Z'^\alpha = (g Z)^\alpha = g^\alpha_{~\beta} ~Z^\beta
\label{tw10}
\eeq
where $g \in SU(4)$.
The generators of infinitesimal $SU(4)$ transformations are thus given by
\beq
L^\alpha_{~\beta} = Z^\alpha {\del \over \del Z^\beta} - {1\over 4}\delta^\alpha_{~\beta}
\left( Z^\gamma {\del \over \del Z^\gamma}\right) 
\label{tw11}
\eeq
This may be split into different types of transformations as follows.
\beqar
J_{AB}&=& U_A {\del \over \del U^B} + U_B {\del \over \del U^A}
\hskip .4in SU_L(2)\nonumber\\
J_{{\dot A}{\dot B}} &=& U_{\dot A} {\del \over \del U^{\dot B}} + U_{\dot B} {\del \over \del U^{\dot A}} \hskip .4in SU_R(2)\nonumber\\
P^{A{\dot A}} &=& U^A {\del \over \del W_{{\dot A}}}
\hskip 1in{\rm Translation}\label{tw12}\\
K_{{\dot A} A} &=& W_{{\dot A}} {\del \over \del U^A} \hskip 1in  {\rm Special
~conformal}\nonumber\\
&&\hskip  1.55in{\rm transformation}\nonumber\\
D&=& W_{{\dot A}}{\del \over \del W_{{\dot A}}} - U^A {\del \over \del U^A}
\hskip .3in{\rm Dilatation}\nonumber
\eeqar
where we have also indicated the interpretation of each type of generators. We see that the $SU(4)$ group is indeed the Euclidean conformal group; it is realized in a linear
and homogeneous fashion
on the twistor variables $Z^\alpha$. At the level of the purely holomorphic transformations, one can also choose the Minkowski signature, where upon the transformations given above become conformal transformations in Minkowski space, forming the group
$SU(2,2)$.
\subsection{Supertwistors}

On can generalize  the twistor space to an ${\cal N}$- extended supertwistor space by adding fermionic or Grassman coordinates $\xi_i$, $i =1, 2, ..., {\cal N}$; thus supertwistor space is parametrized by
$(Z^\alpha, \xi_i )$, with the identification
$Z^\alpha \sim \lambda Z^\alpha$, $\xi_i \sim \lambda \xi_i$, where $\lambda$ is
any nonzero complex number \cite{ferber}. $\lambda$ is bosonic, so only one of the bosonic dimensions is removed by this identification. Thus the supertwistor space is ${\bf CP}^{3\vert {\cal N}}$.

The case of ${\cal N} =4$ is special. In this case, one can form a top-rank holomorphic
form on the supertwistor space; it is given by
\beq
\Omega = {1\over 4!}\epsilon_{\alpha\beta\gamma\delta} Z^\alpha dZ^\beta dZ^\gamma
dZ^\delta ~d\xi_1 d\xi_2 d\xi_3 d\xi_4
\label{tw13}
\eeq
Notice that the bosonic part gets a factor of $\lambda^4$ under the transformation
$Z^\alpha \rightarrow \lambda Z^\alpha$, $\xi_i \sim \lambda \xi_i$, while the fermionic part has a factor of
$\lambda^{-4}$. $\Omega$ is thus invariant under such scalings and becomes a differential form on the supermanifold ${\bf CP}^{3\vert 4}$.

At this point, it is worth recalling the Calabi-Yau theorem \cite{GSW}.
\begin{quotation}
\noindent
{\it Theorem}: For a given complex structure and K\"ahler class on a K\"ahler manifold, there exists a unique Ricci flat metric  if and only if the first Chern class of the manifold vanishes or if and only if there is a globally defined  top-rank holomorphic form on the manifold.
\end{quotation}
This is for an ordinary manifold. For the supersymmetric case, we will define a Calabi-Yau supermanifold as one which admits a globally defined top-rank holomorphic 
differential form \cite{sethi}.
Whether such spaces admit a generalization of the Calabi-Yau theorem is not known.
(For ${\cal N}=1$ spaces, a counterexample is known. However, super-Ricci
flatness may follow from the vanishing of the first Chern class for ${\cal N}\geq 2$
 \cite{rocek}.)

\subsection{Lines in twistor space}

Holomorphic lines in twistor space will turn out to be important for the construction of Yang-Mills amplitudes.
First we will consider a holomorphic straight line, or a curve of degree one, in twistor space, giving the generalization
to supertwistor space later. Since ${\bf CP}^3$ has three complex dimensions,
we need two complex conditions to reduce to a line in twistor space.
Thus we can specify a line in twistor space as the solution set of the equations
\beq
A_\alpha Z^\alpha =0, \hskip .5in B_\alpha Z^\alpha =0
\label{tw14}
\eeq
where $A_\alpha, ~B_\alpha$ are constant twistors which specify the placement of the line in twistor space.  These equations can be combined as
\beq
a_A^i U^A ~+~ b_{\dot A}^i W^{\dot A} =0
\label{tw15}
\eeq
where $A_\alpha = (a_A^1, b_{\dot A}^1) ,~ B_\alpha =( a_A^2, b_{\dot A}^2)$.
$a, ~b$ can be considered as $(2\times 2)$-matrices; $\det a$ and $\det b$ may both be nonzero, but both cannot be zero simultaneously, since equations (\ref{tw14}) are then not sufficient to reduce to a line.
We will take $\det b \neq 0$ in the following. (The arguments presented will go through
with appropriate relabelings if $\det b =0$, but $\det a\neq 0$.)
In this case, $b$ is invertible and we can solve the equations (\ref{tw15}) by
\beqar
W_{\dot A} &=& - (b^{-1} a )_{{\dot A}A}U^A \nonumber\\
&\equiv& x_{{\dot A}A} U^A\label{tw16}
\eeqar
This shows that the condition (\ref{tw4}) identifying the spacetime coordinates may be taken as defining a line in twistor space. In fact, here, $x_{{\dot A}A}$ specify the placement
and orientation of the line in twistor space, in other words, they are the moduli of the line.
We see that the moduli space of straight lines (degree-one curves) in twistor
space is spacetime.

There is another way to write the equation (\ref{tw16}). Recall that a line
in a real space $M$ can be defined as a mapping of the interval $[0,1]$ into the space
$M$, $L: [0,1]\rightarrow M$. We can do a similar construction for the complex case.
We will define an abstract
${\bf CP}^1$ space by a two-spinor $u^a$ with the identification
$u^a \sim \rho u^a$ for any nonzero complex number $\rho$.
Then we can regard a holomorphic line in twistor space as
a map ${\bf CP}^1 \rightarrow {\bf CP}^3$, realized explicitly as
\beq
U^A = (a^{-1})^A_a u^a, \hskip .5in W_{\dot A} = (b^{-1})_{{\dot A}a} u^a
\label{tw17}
\eeq
One can do $SL(2,{\bf C})$ transformations on the coordinates $u^a$ of
${\bf CP}^1$; using this freedom, we can set $a=1$, or equivalently,
\beq
U^A = u^A, \hskip .6in W_{\dot A}= x_{{\dot A} A}u^A
\label{tw18}
\eeq
This is identical to (\ref{tw16}).

The generalization of this to supertwistor space is now obvious. We will consider
a map of ${\bf CP}^1$ to the supertwistor space ${\bf CP}^{3\vert 4}$, given explicitly
by
\beqar
&&U^A = u^A, \hskip .6in W_{\dot A}= x_{{\dot A} A}u^A\nonumber\\
&&\xi^\alpha = \theta^\alpha_A u^A\label{tw19}
\eeqar
We have the fermionic moduli $\theta^\alpha_A$ in addition to the bosonic ones
$x_{{\dot A}A}$.

The construction of curves of higher degree can be done along similar lines.
A curve of degree $d$ is given by
\beqar
Z^\alpha &=& \sum_{\{a\}} a^\alpha_{a_1 a_2 \cdots a_d} ~ u^{a_1} u^{ a_2} \cdots u^{ a_d}
\nonumber\\
\xi^\alpha &=& \sum_{\{ a\}} \gamma^\alpha_{a_1 a_2 \cdots a_d} ~ u^{a_1} u^{ a_2} \cdots u^{ a_d}\label{tw20}
\eeqar
The coefficients $a^\alpha_{a_1 a_2 \cdots a_d}$, $\gamma^\alpha_{a_1 a_2 \cdots a_d}$ give the moduli of the curve. One can use $SL(2,{\bf C})$ to set three of the coefficients to
fixed values.

Given that each index $a$ takes values $1, 2$, and the fact that the coefficients are
symmetric in $a_1, a_2, ..., a_n$, we see that there are $4 (d+1) $ bosonic and fermionic
coefficients. The identification of
$u^a $ and $\rho u^a$ and $Z\sim \lambda Z, ~ \xi \sim \lambda \xi$
tells us that we can remove an overall scale degree of freedom. 
In other words, for the moduli, we have the identification,
\beqar
&&a^\alpha_{a_1 a_2 \cdots a_d}\sim \lambda ~
a^\alpha_{a_1 a_2 \cdots a_d}\nonumber\\
&&\gamma^\alpha_{a_1 a_2 \cdots a_d}\sim \lambda~
\gamma^\alpha_{a_1 a_2 \cdots a_d}
\label{tw21}
\eeqar
for $\lambda \in {\bf C}- \{ 0\}$. Thus the moduli space of the curves may be taken as
${\bf CP}^{4d+3\vert 4d+4}$.
For the expressions of interest, as we shall see later, there is an overall $SL(2, {\bf C})$
invariance, and hence three of the bosonic parameters can be fixed to
arbitrarily chosen values.
   
\section{Yang-Mills amplitudes and twistors}
\label{sec:7}
\subsection{Why twistors are useful}

In this section, we will start our discussion of the twistor approach to amplitudes in Yang-Mills theory \cite{nair, witten}, or multigluon scattering amplitudes, as they are often referred to.

We begin with the question of why the calculation of multigluon amplitudes are interesting.
One of the motivations in seeking a twistor string theory was to obtain a weak coupling version of the standard duality between string theory on anti-de Sitter space and ${\cal N}=4$ supersymmetric Yang-Mills theory \cite{witten}.
However, developments in the subject over the last year or so have focused on, and yielded, many interesting results on the calculation of the scattering amplitudes themselves, so we shall 
concentrate on this aspect of twistors
\cite{CSK}. Scattering amplitudes in any gauge theory are not only interesting in a general sense of helping to clarify a complicated interacting theory, but also there is a genuine need for them
from a very practical point of view. This point can be illustrated by taking the 
experimental determination of the strong coupling constant as an example.
We quote three values from the Particle Data Group based on three different processes.
\beq
\begin{array}{r l l l}
\alpha_s &= 0.116~&~+~0.003 ~(expt.)&\pm ~0.003 ~(theory)\\
&&~-~0.005&\\
&&&\\
&=0.120~&~\pm ~0.002~ (expt.)&\pm ~0.004~ (theory)\\
&&&\\
&=0.1224&~\pm ~0.002~ (expt.)&\pm ~0.005~ (theory)
\end{array}
\label{YM1}
\eeq
These values are for the momentum scale corresponding to the mas of the $Z$-boson, namely, $\alpha_s (M_Z)$; they are based on the Bjorken spin sum rule, jet rates in $e$-$p$ collisions and the photoproduction of two or more jets, respectively. Notice that the theoretical uncertainty is comparable to, or exceeds, the experimental errors. The major part of this
comes from lack of theoretical calculations (to the order required) for the processes from which this value is extracted.
Small as it may seem, this uncertainty can affect the hadronic background analysis at the Large Hadron Collider (currently being built at CERN), for instance. The relative signal strength for processes of interest, such as the search for the Higgs particle, can be improved if this uncertainty is reduced. This can also affect the estimate of the grand unification scale and theoretical issues related to it.

One could then ask the question: Since we know the basic vertices involved, and these are ultimately perturbative calculations, why not just do the calculations, to whatever order is required? Unfortunately, the direct calculation of the amplitudes is very, very difficult since there are large numbers, of the order of millions, of Feynman diagrams involved. (It is easy to see that the number of diagrams involved increases worse than factorially as the number of external lines increases.) Twistors provide a way to improve the situation.

A natural next question is then: What can twistors do, what has been accomplished so far?
The progress so far may be summarized as follows.
\begin{enumerate}
\item It has been possible to write down a formula for all the tree level amplitudes in
${\cal N}=4$ supersymmetric Yang-Mills theory
\cite{list1, list2}. Being at the tree level, this formula
applies to the tree amplitudes of the nonsupersymmetric Yang-Mills theory as well.
The formula reduces the calculation to the evaluation of the zeros of a number of polynomial equations and the evaluation of an integral. A certain analogy with instantons might be helpful in  explaining the nature of this formula. To find the instanton field configurations, one must solve the  self-duality conditions which are a set of coupled first-order differential equations.
However, the ADHM procedure reduces this to an algebraic problem, namely, of solving a set of matrix equations, which lead to the construction of the appropriate holomorphic vector bundles. This algebraic problem is still difficult for large instanton numbers, nevertheless, an algebraicization has been achieved. In a similar way, the formula for the tree amplitudes replaces the evaluation of large numbers of Feynman diagrams, or an equivalent functional integral, by an ordinary integral whose evaluation requires the solution of some polynomial equations. This may still  be difficult for cases with large numbers of negative helicity gluons; nevertheless, it is a dramatic simplification.

\item At the one-loop level, a similar formula has been obtained for all the so-called maximally helicity violating (MHV) amplitudes \cite{oneloop}. A number of results for next-to-MHV amplitudes have been obtained \cite{oneloop2}.

\item A set of new diagrammatic rules based on the MHV vertices has been developed
\cite{CSW}.
Also, new types of recursion rules have been developed
\cite{recursion}. These are promising new directions for perturbative 
analysis of a field theory. 
\item Twistor inspired techniques have been used for some processes involving
massive particles, particularly for the electroweak calculations \cite{higgs}.
\end{enumerate}

\subsection{The MHV amplitudes}

We will begin with a discussion of the maximally helicity violating (MHV) amplitudes.
These refer to $n$ gluon scattering amplitudes with $n\!-\!2$ gluons of positive helicity and $2$ gluons of negative helicity. 
With $n$ gluons, $n\!-\!2$ is the maximum number of positive
helicity possible by conservation laws and such amplitudes are
often refered to as the maximally helicity-violating (MHV)
amplitudes.The analysis of these amplitudes will lead the way to the generalization for all amplitudes.

The gluons are massless and have momenta $p_\mu$ which obey
the condition $p^2 =0$; i.e., $p_\mu$ is a null vector.
Using the identity and the Pauli matrices, we can write the four-vector
$p_\mu$ as a $2\times 2$-matrix
\beq
p^A_{~\dotA} = (\sigma^\mu)^A_{~\dotA}
~p_\mu = \left( \begin{matrix} p_0 +p_3&~ p_1 -ip_2\\
p_1 +ip_2&~ p_0 -p_3\\
\end{matrix}\right)
\label{YM2}
\eeq
This matrix has a zero eigenvalue and,
using this fact, we can see that it can be written as
\beq
p^A_{~\dotA} = \pi^A \bpi_{\dotA}
\label{YM3}
\eeq
There is a phase ambiguity in the definition
of $\pi , ~\bpi$; $\pi' = e^{i\theta}\pi,~ \bpi' = e^{-i\theta}\bpi$
give the same momentum vector $p_\mu$. Thus, physical results should be independent of
this phase transformation.
For a particular choice of this phase, an explicit
realization of $\pi,~\bpi$ is given by
\beq
\pi ={1\over \sqrt{p_0-p_3}}\left( \begin{matrix} {p_1 -ip_2}\\
{p_0-p_3}\\
\end{matrix}\right),
\hskip .3in
\bpi ={1\over \sqrt{p_0-p_3}}\left( \begin{matrix} {p_1 +ip_2}\\
{p_0-p_3}\\
\end{matrix}\right)
\label{YM4}
\eeq
The fact that $p_\mu$ is real gives a condition between
$\pi$ and $\bpi$, which may be taken as $\bpi_{\dotA} =(\pi^A)^*$.
We see that, for the momentum for each massless
particle, we can associate a spinor momentum $\pi$. 

There is a natural action of the Lorentz group $SL(2,{\bf C})$
on the dotted and undotted indices, given by
\beq
\pi^A \rightarrow \pi'^A =
(g \pi )^A, \hskip .3in
\bpi_{\dotA} \rightarrow \bpi'_{\dotA}
= (g^{*}\bpi )_{\dotA}
\label{YM5}
\eeq
where $g \in SL(2, {\bf C})$.
The scalar product which preserves this
symmetry is given by $\la 12\ra= \pi_1\cdot \pi_2 = \epsilon_{AB} \pi^A_1
\pi^B_2$, $[12]= \epsilon^{\dotA\dotB} \bpi_{1\dotA}
\bpi_{2\dotB}$.  
At the level of vectors, this corresponds to the Minkowski product;
i.e., $\eta^{\mu\nu}p_{1\mu} p_{2\nu} = p_1 \cdot p_2 = \la 12\ra [12]$.
 Because of the $\epsilon$-tensor, $\la 11\ra=0$
and the factorization (\ref{YM3}) is consistent with $p^2 =0$.
(More generally, $\la 12\ra =0$ and $[12]=0$ if $\pi_1$ and
$\pi_2$ are proportional to each other.) The scattering amplitudes can be simplified considerably when expressed in terms of these invariant spinor products.
We will also define raising and lowering of the spinorial indices
using the $\epsilon$-tensor.

It is also useful to specify the polarization states of the gluons by helicity.
The polarization vector $\epsilon_\mu$ may then be written as
\beq
\epsilon_\mu \rightarrow \epsilon^A_{~\dot A}= (\sigma^\mu)^A_{\dot A}~ \epsilon_\mu
= \left\{ \begin{matrix}{\lambda^A \bpi_{\dot A}/ \pi\cdot \lambda}
&~\hskip .1in+1~{\rm helicity}\\
&\\
{\pi^A {\bar\lambda}_{\dot A} / \bpi\cdot {\bar \lambda}}&~\hskip .1in -1 ~{\rm helicity}\\
\end{matrix}
\right.
\label{YM6}
\eeq
The spinor $(\lambda^A, ~{\bar\lambda}_{\dot A})$ characterizes the choice of helicity.

We can now state the MHV amplitude for scattering of $n$ gluons, originally obtained by
Parke and Taylor \cite{pt}. They carried out the explicit calculation of Feynman diagrams, for small values of $n$, using some supersymmetry tricks for simplifications. Based on this, they guessed the general form of the amplitude; this guess was proved by Berends and Giele by using recursion rules for scattering amplitudes \cite{pt}.
The results are the following.
\beqar
{\cal A}( 1^{a_1}_+ , 2^{a_2}_+, 3^{a_3}_+, \cdots , n^{a_n}_+ )&=& 0\nonumber\\
{\cal A} ( 1^{a_1}_- , 2^{a_2}_+, 3^{a_3}_+, \cdots , n^{a_n}_+ )&=& 0\nonumber\\
{\cal A}( 1^{a_1}_- , 2^{a_2}_-, 3^{a_3}_+, \cdots , n^{a_n}_+ )&=& 
i g^{n-2} 
(2\pi )^4 \delta (p_1+p_2+...+p_n) ~{\cal M}\label{YM7}\\
&&\hskip .4in +{\rm noncyclic ~permutations}\nonumber\\
{\cal M}( 1^{a_1}_- , 2^{a_2}_-, 3^{a_3}_+, \cdots , n^{a_n}_+ ) &=& \la 12\ra^4 ~{ \Tr (t^{a_1} t^{a_2} \cdots t^{a_n}) \over \la 12\ra \la 23\ra \cdots 
\la n-1~n\ra \la n 1\ra}\nonumber
\eeqar
The first nonvanishing amplitude with the maximum difference of helicities has
two negative helicity gluons and 
$n\!-\!2$ positive helicity gluons. This is what is usually called the MHV amplitude.
In equation (\ref{YM7}), $g$ is the gauge coupling constant. Notice that the amplitude
${\cal M}$ is cyclically symmetric in all the particle labels
except for the prefactor $\la 12\ra^4$. The latter refers to the momenta of the two negative helicity gluons. The summation over the noncyclic permutations makes the full amplitude symmetric in the gluon labels. We have taken all gluons as incoming. One can use 
the standard crossing symmetry to write down the corresponding amplitudes, with appropriate change of helicities, if some of the gluons are outgoing.

We will now carry out three steps of simplification of this result to bring out the twistor connection. 
\vskip .1in\noindent
{$\underline{The ~first ~step: ~The ~chiral ~Dirac ~determinant ~on ~{\bf CP}^1}$}
\vskip .1in
Consider the functional determinant of the Dirac operator of a chiral fermion coupled to a gauge field $A_\bz$
in two dimensions. By writing $\log \det D_\bz = \Tr \log D_\bz$ and expanding the
logarithm, we find
\beqar
\Tr \log D_\bz &=& \Tr \log (\del_\bz +A_\bz )\nonumber\\
&=& \Tr \log \left( 1 + {1\over \del_\bz} A_\bz \right) ~~~+~ {\rm constant}\nonumber\\
&=& \sum_n \int {d^2x_1 \over \pi}{d^2x_2 \over \pi} \cdots
{(-1)^{n+1} \over n} { \Tr [ A_\bz (1) A_\bz (2) \cdots A_\bz (n) ] \over 
z_{12} z_{23}\cdots z_{n-1 ~n} z_{n1} }\nonumber\\
\label{YM8}
\eeqar
where $z_{12} = z_1 - z_2$, etc. In writing this formula we have used the result
\beq
\left( {1\over \del_\bz }\right)_{12}= {1\over \pi (z_1 - z_2)}
\label{YM9}
\eeq

We can regard the $z$'s as local coordinates on ${\bf CP}^1$. Recall that ${\bf CP}^1$ is defined by two complex variables $\alpha$ and $\beta$, which may be regarded as a two-spinor $u^a$, $u^1 =\alpha, ~u^2 =\beta$, with the identification $u^a \sim \rho u^a$, $\rho \in {\bf C} \!-\! \{ 0 \}$. On the coordinate patch with $\alpha \neq 0$, we take
$z = \beta /\alpha$ as the local coordinate.
We can then write
\beqar
z_1 - z_2 &=& {\beta_1 \over \alpha_1}- {\beta_2\over \alpha_2} = 
{\beta_1 \alpha_2 - \beta_2 \alpha_1\over \alpha_1 \alpha_2}\nonumber\\
&=& {\epsilon_{ab} u^a_1 u^b_2 \over \alpha_1 \alpha_2} = {u_1 \cdot u_2 \over
\alpha_1\alpha_2}
\label{YM10}
\eeqar
Further, if we define $ \alpha^2 A\bz = {\bar{\cal A}}$, equation (\ref{YM8}) becomes
\beq
\Tr \log D_\bz = - \sum {1\over n} \int {\Tr[ {\bar{\cal A}}(1) {\bar{\cal A}}(2) \cdots
{\bar{\cal A}}(n) ]\over (u_1\cdot u_2) (u_2\cdot u_3) \cdots (u_n \cdot u_1)}
\label{YM11}
\eeq
Notice that if $u^a$ is replaced by the spinor momentum  $\pi^A$, the denominator is exactly
what appears in (\ref{YM7}). The factor of $1/n$ gets cancelled out because (\ref{YM11})  generates all permutations which gives $n$ times the sum over all noncyclic permutations.
\vskip .1in\noindent
{$\underline{The ~second ~step: ~The ~helicity ~ factors}$}
\vskip .1in
The denominator for the MHV amplitude can be related to the chiral Dirac determinant
as above. The factor $\la 12\ra^4$ can also be obtained if we introduce supersymmetry. We take up this second step of simplification now.

The transformation (\ref{YM5})
shows that the Lorentz generator for the $\pi$'s is given by
\beq
J_{AB} = {1\over 2}\left( \pi_A {\del \over \del \pi^B}
+ \pi_B {\del\over \del \pi^A}\right)
\label{YM12}
\eeq
where $\pi_A = \epsilon_{AB} \pi^B$.
The spin operator is given by $S_\mu \sim \epsilon_{\mu\nu\alpha\beta}
J^{\nu\alpha}p^\beta$, where $J^{\mu\nu}$ is the full Lorentz generator.
This works out to
$S_{A \dotA}
= J^A_{B} \pi^{B}\bpi_\dotA = -p^A_{~\dotA}~ s$, identifying the helicity
as
\beq
s = -{1\over 2} \pi^A {\del \over \del \pi^A}
\label{YM13}
\eeq
Thus $s$ is, up to a minus sign, half the degree of homogeneity in the $\pi$'s.
If we start with a positive helicity gluon, which would correspond to two
negative powers of the corresponding spinor momentum, then we should expect
an additional four factors of $\pi$ for a negative helicity gluon.
Notice that there are two factors of spinor momenta in the denominator of the scattering amplitude (\ref{YM7}) for each positive helicity gluon; for the two negative helicity gluons,
because of the
extra factor of $\la 12\ra^4$, the net result is two positive powers of $\pi$.

We now notice that if we have an anticommuting spinor
$\theta_A$, $\int d^2 \theta ~\theta_A \theta_B = \epsilon_{AB}$,
so that
\beq
\int d^2\theta ~(\pi\theta ) (\pi'\theta ) = \int d^2\theta ~(\pi^A\theta_A) (\pi'^{B}\theta_B ) 
= \pi \cdot \pi'\label{YM14}
\eeq
We see that an ${\cal N}=4$ theory is what we need to get
four such factors, so as to get a term like $\la 12\ra^4$. Therefore we define an ${\cal N}=4$ superfield
\beq
\bA^a (\pi,\bpi ) =
a^a_{+} + \xi^\alpha a^a_\alpha + {1\over 2} \xi^\alpha \xi^\beta a^a_{\alpha\beta} +{1\over 3!}
\xi^\alpha \xi^\beta \xi^\gamma \epsilon_{\alpha\beta\gamma\delta} {\bar a}^{a\delta} +\xi^1 \xi^2 \xi^3 \xi^4 a^a_{-}
\label{YM15}
\eeq
where $\xi^\alpha = (\pi\theta )^\alpha =
\pi^A \theta^\alpha_A$, $\alpha=1, 2, 3, 4$.
We can interpret $a^a_+$ as the classical version
of the annihilation operator for a positive
helicity gluon (whose gauge charge is specified by the Lie algebra index $a$), $a^a_-$ as the annihilation operator for a negative
helicity gluon; $a^a_\alpha , {\bar a}^{a\alpha}$
correspond to four spin-${1\over 2}$ particles and $a^a_{\alpha\beta}$ 
correspond to six spin-zero
particles. This is exactly the particle content of ${\cal N}=4$ Yang-Mills theory.

We now choose the gauge potential in (\ref{YM11}) to be given by
\beq
{\bar {\cal A}} = g t^a \bA^a \exp (ip\cdot x )
\label{YM16}
\eeq
Using this in the chiral Dirac determinant (\ref{YM11}), we construct the expression
\beq
\Gamma [a] = {1\over g^2}\int d^8\theta d^4x~ \Tr \log D_\bz \Biggr]_{u^a\rightarrow \pi^A}
\label{YM17}
\eeq 
It is then clear that the MHV amplitude can be written as
\beqar
{\cal A}( 1^{a_1}_- , 2^{a_2}_-, 3^{a_3}_+, \cdots , n^{a_n}_+ )&\!\!=\!\!& 
i\left[{\delta \over \delta a^{a_1}_{-} (p_1)}{\delta \over \delta a^{a_2}_{-} (p_2)}
{\delta \over \delta a^{a_3}_{+} (p_3)}\cdots {\delta \over \delta a^{a_n}_{+} (p_n)}
\Gamma [a] \right]_{a=0}\nonumber\\
\label{YM18}
\eeqar

An alternate representation involves introducing a supersymmetric
version of the factor $\exp (ip\cdot x)$ for the 
${\cal N}=4$ supermultiplet. Consider the function
\beqar
\exp (i \eta \cdot \xi )
= 1 ~+~ i\eta \cdot \xi ~+~ {1\over 2!} i\eta \cdot \xi ~i\eta \cdot \xi
~+~ {1\over 3!}i\eta \cdot \xi ~i\eta \cdot \xi ~i\eta \cdot \xi&&\nonumber\\
~+~{1\over 4!}i\eta \cdot \xi~i\eta \cdot \xi~i\eta \cdot \xi~i\eta \cdot \xi&&
\label{YM19}
\eeqar
where $\eta \cdot \xi = \eta_\alpha \xi^\alpha$. Here $\eta_\alpha$ are four Grassman variables. We may think of them as characterizing the state of the external particles,
specifically, their helicity. (The state is thus specified by the spinor momentum $\pi$ and
$\eta$.)
Since each $\xi$ carries one power of $\pi$, we see that we can associate
the first term with a positive helicity gluon, the last with a negative helicity gluon,
and the others with the superpartners of gluons, accordingly.
We can then take 
\beq
{\bar {\cal A}} = g t^a \phi^a \exp (ip\cdot x +i \eta \cdot \xi )
\label{YM20}
\eeq
The scattering amplitudes are given by $\Gamma [a]$ again, where, to get negative helicity for particles labeled $1$ and $2$ we take the coefficient of the factor
$\eta_{11} \eta_{21} \eta_{31} \eta_{41} \eta_{12} \eta_{22} \eta_{32} \eta_{42} $,
where the first subscript gives the component of $\eta$ and the second refers to the particle.
We should also look at the term with $n$ factors of $\phi^a$ for the $n$-gluon amplitude.
\vskip .1in\noindent
{$\underline{The ~third ~step: ~Reduction ~to ~a ~line ~in ~twistor ~space}$}
\vskip .1in
The results (\ref{YM18}-\ref{YM20}) were known for a long time. The importance of
supertwistor space was also recognized \cite{nair}. 
(Some of the earlier developments, with connections to
the self-dual Yang-Mills theory, etc., can be traced from \cite{others}.)
Notice that with $U^A = u^A$, $W = xu$ and
$\xi$, alongwith the condition $\pi^A = U^A$, we are close to the usual variables of supertwistor space. More recently, Witten
achieved enormous advances in this field by relating this formula to 
twistor string theory
and curves in twistor space \cite{witten}.
To arrive at this generalization, first of all, we notice that the amplitude is holomorphic
in the spinor momenta except for the exponential factor $\exp (ip\cdot x)$.
We can rewrite this factor as follows.
\beqar
\exp (ip\cdot x) &=&  \exp \left( {i\over 2} \bpi^{\dot A} x_{{\dot A}A} \pi^A\right)\nonumber\\
&=& \exp \left( {i \over 2} \bpi^{\dot A} W_{\dot A}\right)\Biggr]_{u^A =\pi^A}
\label{YM21}
\eeqar
where $W_{\dot A} = x_{{\dot A}A} \pi^A$. The strategy is now to
regard $W_{\dot A}$ as a free variable, interpreting the condition
$W_{\dot A} = x_{{\dot A}A} u^A$ as the restriction to a line in twistor space.
We shall also use $U^A = u^A$, which is the other condition defining a line
in twistor space. see (\ref{tw18}).
We can then write
\beq
\int d\sigma ~ \delta \left( {\pi^2 \over \pi^1} - {U^2 \over U^1}\right)
\exp\left( {i\over 2} \bpi^{\dot A} \pi^1 {W_{\dot A} \over U^1}\right)
=\exp( {{i\over 2} \bpi^{\dot A} x_{{\dot A}A}\pi^A}) = \exp (i p\cdot x)  \label{YM22}
\eeq
where $\sigma = u^2 /u^1$, and we have used the restriction to the
line $W_{\dot A} = x_{{\dot A}A} u^A$, $U^A = u^A$.
The integration is along a line which contains the support of the $\delta$-function.

We can also treat $\xi$ as an independent variable, 
interpreting the condition $\xi^\alpha = \theta^\alpha_A u^A$ as part of the line in
supertwistor space, as in (\ref{tw19}). The amplitude for $n$ particle scattering, with
particle momenta labeled by $\pi^A_i,~\bpi^{\dot A}_i$ and helicity factors
$\eta_{\alpha i}$, can then be written as
\beqar
{\cal A}
&=& ig^{n-2}\int d^4x d^8\theta~ \int d\sigma_1 \cdots d\sigma_n
{\Tr (t^{a_1} \cdots t^{a_n})\over {(\sigma_1 -\sigma_2)(\sigma_2 -\sigma_3)
\cdots (\sigma_n -\sigma_1)}}\nonumber\\
&&\hskip .3in
\times\prod_i \delta \left( {\pi_i^2 \over \pi_i^1} - {U^2(\sigma_i) \over U^1(\sigma_i)}\right) \exp \left( {i\over 2} \bpi_i^{\dot A} \pi_i^1 {W_{\dot A}(\sigma_i) \over U^1(\sigma_i)}
+i\pi^1_i \eta_{\alpha i}{\xi^\alpha (\sigma_i)\over U^1(\sigma_i)}\right)\nonumber\\
&&\hskip .3in + {\rm noncyclic ~permutations}
\label{YM23}
\eeqar
where the functions $W_{\dot A}, ~U^A, ~\xi^\alpha$ are given by
\beqar
&&U^A = u^A, \hskip .6in W_{\dot A}= x_{{\dot A} A}u^A\nonumber\\
&&\xi^\alpha = \theta^\alpha_A u^A\label{YM24}
\eeqar
exactly as in (\ref{tw19}). The variable $\sigma$ is given by
$\sigma = u^2/u^1$. Notice that the overall factor of $u^1$ 
in $W_{\dot A}, ~U^A, ~\xi^\alpha$, cancels out in
the formula (\ref{YM23}).

\subsection{Generalization to other helicities}

The amplitude, in the form given in (\ref{YM23}), shows a number of interesting properties.
First of all, the amplitude is
entirely holomorphic in the twistor variables $Z^\alpha = (W_{\dot A}, U^A)$, $\xi^\alpha$.
It is also holomorphic in the variable $\sigma$ or $u^a$.
(It is not holomorphic in $\pi$ since there is $\bpi$ in the exponentials, but this is immaterial for our arguments given below.)
Secondly, the amplitude is invariant under the scalings
$Z^\alpha \rightarrow \lambda Z^\alpha$, $\xi^\alpha \rightarrow \lambda \xi^\alpha$,
so that it is a properly defined function on some neighborhood in the supertwistor space.
Further, the amplitude has support only on a curve of degree one in supertwistor space
given by (\ref{YM24}). 
The moduli of this curve are given by $x_{{\dot A}A}$ and $\theta^\alpha_A$;
there is integration over all these in the amplitude.

We may interpret this as follows. We consider a holomorphic map ${\bf CP}^1 \rightarrow
{\bf CP}^{3|4}$ which is of degree one.  We pick $n$ points
$\sigma_1, \sigma_2,\cdots, \sigma_n$ and then evaluate
the integral in (\ref{YM23}) over all $\sigma$'s and the moduli of the
chosen curve.

The generalization of the formula suggested by Witten is to use curves
of higher degree \cite{witten}. In fact Witten argued, based on twistor string theory, that
one should consider curves of degree $d$ and genus $g$, with
\beq
d = q-1 +l , \hskip .5in g \leq l
\label{YM25}
\eeq
for $l$-loop Yang-Mills amplitudes with $q$ gluons of negative helicity.
This generalization has been checked for various cases as mentioned before. For the
tree amplitudes, the generalized formula reads \cite{witten, list1}
\beqar
{\cal A}
&=&ig^{n-2} \int d\mu ~ \int d\sigma_1 \cdots d\sigma_n
{\Tr (t^{a_1} \cdots t^{a_n})\over {(\sigma_1 -\sigma_2)(\sigma_2 -\sigma_3)
\cdots (\sigma_n -\sigma_1)}}\nonumber\\
&&\hskip .3in
\times\prod_i \delta \left( {\pi_i^2 \over \pi_i^1} - {U^2(\sigma_i) \over U^1(\sigma_i)}\right) \exp \left( {i\over 2} \bpi_i^{\dot A} \pi_i^1 {W_{\dot A}(\sigma_i) \over U^1(\sigma_i)}
+i\pi^1_i \eta_{\alpha i}{\xi^\alpha (\sigma_i)\over U^1(\sigma_i)}\right)\nonumber\\
&&\hskip .3in + {\rm noncyclic ~permutations}
\label{YM26}
\eeqar
where the curves of degree $d$ are
\beqar
&&W_{\dot A} (\sigma )= (u^1)^d \sum_0^d b_{{\dot A}k} \sigma^k, \hskip .4in
U^A (\sigma )= (u^1)^d \sum_0^d a^A_k \sigma^k\nonumber\\
&& \xi^\alpha (\sigma ) = (u^1)^d \sum_0^d \gamma^\alpha_k \sigma^k
\label{YM27}
\eeqar
This is exactly as in (\ref{tw20}). The measure of integration for the moduli
in (\ref{YM26}) is given by
\beq
d\mu = {d^{2d+2}a~~ d^{2d+2}b~ ~d^{4d+4} \gamma \over 
vol [GL(2,{\bf C})]}
\label{YM28}
\eeq
The division by the volume of $GL(2,{\bf C})$ arises as follows.
There is an overall scale invariance for the integrand in (\ref{YM26}),
which means that we can remove one complex scale factor, corresponding to the
moduli space being ${\bf CP}^{4d+3|4d+4}$.
The integrand is also holomorphic in $\sigma$ and so has invariance under
the $SL(2, {\bf C})$ transformations
$u^a \rightarrow u'^a = (g u)^a$
where $g$ is a $(2\times 2)$-matrix with unit determinant, or an element of
$SL(2, {\bf C})$. We must remove this factor to get an integral which does not
diverge.

The actual evaluation of the integral can still be quite involved.
One has to identify the zeros of the functions $U^2(\sigma )/U^1(\sigma )$
to integrate over the $\delta$-functions. This can be difficult to do
explicitly for arbitrary values of the moduli. This is then followed by the
integration over the moduli. Nevertheless, the formulae (\ref{YM26}, \ref{YM27})
constitute a significant achievement. They reduce the problem of amplitude calculations
in the gauge theory to an ordinary, multidimensional integral. As mentioned at the beginning of this section,
this reduction has a status somewhat similar to what
the ADHM construction has achieved for instantons.

Another important qualification about the formula (\ref{YM26})
is that the integrals have to be defined by a continuation to real
variables.
The spacetime signature has to be chosen to be $(--++)$
to be compatible with this. One has to carry out the analytic continuation
after the integrals are done.

The justification for the generalization embodied in (\ref{YM26})
comes from twistor string theory.
We shall now briefly review this connection following
Witten's construction of twistor string theory \cite{witten}; there is an alternative
string theory proposed by Berkovits which can also be used \cite{berkovits}.
Some of the structure of the latter will be used in section \ref{sec:9}.

\section{Twistor string theory}
\label{sec:8}
As mentioned in section \ref{sec:6}, supertwistor space ${\bf CP}^{3|4}$
is a Calabi-Yau space. This allows the construction of a topological $B$-model
with ${\bf CP}^{3|4}$ as the target space. In this theory, one considers open strings which end
on $D5$-branes, with the condition $\bar\xi =0$. The gauge fields which characterize the
dynamics of the ends of the open strings is then a potential
${\bar{\cal A}}(Z, {\bar Z}, \xi )$ which can  be checked to have the same content
as the ${\cal N}=4$ gauge theory.  An effective action for the topological
sector can be written down; it is given by
\beq
{\cal I} = {1\over 2} \int_Y \Omega \wedge \Tr ( {\bar{\cal A}}~ \bdel ~{\bar{\cal A}}
+{2\over 3} {\bar{\cal A}}^3 )
\label{tst1}
\eeq
Here $Y$ is a submanifold of ${\bf CP}^{3|4}$
with $\bar\xi =0$. To linear order in the fields, the equations of motion
for (\ref{tst1}) correspond to $\bdel {\bar{\cal A}} =0$. Thus, the fields
are holomorphic in the $Z$'s and, via the Penrose transform, they correspond to massless
fields in ordinary spacetime.
Including the nonlinear terms, one can ask for an action in terms of the
fields in spacetime which is equivalent to (\ref{tst1}).
This is given by
\beq
{\cal I} = \int \Tr \left[ G^{AB} F_{AB}  + {\bar\chi}^{A \alpha}
D_{A{\dot A}} \chi^{\dot A}_\alpha +\cdots \right]
\label{tst2}
\eeq
$G^{AB}$ is a self-dual field, $F_{AB}$ of the self-dual part of the
usual field strength $F_{\mu\nu}$ of an ordinary gauge potential,
$\chi, {\bar\chi}$ are fermionic fields, etc.
In terms of helicities $G^{AB}$ corresponds to $-1$, the nonvanishing field strength
$F_{{\dot A} {\dot B}}$ corresponds to $+1$ and so on.
The action (\ref{tst2}) cannot generate amplitudes with arbitrary number of
negative helicity gluons; it is not the (super) Yang-Mills action either.
The usual Yang-Mills term can be generated by the effect of a $D1$-instanton,
which can lead to a term of the form
${1\over 2} \int \epsilon G^2$, where $\epsilon$ is related to the action for the instanton. Integrating out the $G$-field, we then get a Yang-Mills term of the form
$\int F^2/4g^2 $ with a Yang-Mills coupling constant $g^2 \sim \epsilon$. 
A term with $q$ factors of $G$, corresponding to $q$ particles of helicity
$-1$, will require $q-1$ powers of $\epsilon$. This corresponds to instanton number
$q-1=d$. Such $D1$-instantons are described by holomorphic curves of degree
$d$. This is in agreement with the formula (\ref{YM25}).
This is the basic argument, schematically, why we should expect curves of degree $d$ to
lead to tree-level amplitudes in the ${\cal N}=4$ Yang-Mills theory.

We have not discussed amplitudes at the loop level yet. 
A similar approach with curves of genus one
has been verified at the one-loop level \cite{oneloop}.
However, the general formula in terms of higher genus curves has not been very useful
for actual computations.
This is partially due to the complexity of the formula. A more relevant reason is the
emergence of new rules to calculate both tree level and loop level amplitudes
using a sewing procedure with the basic MHV amplitudes as the vertices \cite{CSW}.
The MHV amplitudes have to be continued off-shell for this reason. An off-shell extension
has been proposed and used in \cite{CSW}.

The one-loop amplitudes which emerge naturally are the amplitudes for
the ${\cal N}=4$ Yang-Mills theory. 
If the external (incoming and outgoing) particles are gluons, 
the superpartners can only occur in loops. As a result, at
the tree-level, one can get
the amplitudes in the pure Yang-Mills theory with no supersymmetry by restricting the
external lines to be gluons. But at the one-loop level all superpartners can contribute.
While the one-loop amplitudes for the ${\cal N}=4$ Yang-Mills theory are interesting
in their own right, the corresponding amplitudes in the nonsupersymmetric theory
are of even greater interest, since they pertain to processes which are
experimentally accessible. It is not trivial to extract the nonsupersymmetric amplitudes 
from the ${\cal N}=4$ theory. One approach is to subtract out the contributions of the superpartners. An alternative is to build up the one-loop amplitude
from the unitarity relation, using tree amplitudes. In principle, this can only yield the imaginary part of the one-loop amplitude. (One could attempt to construct the real part via dispersion relations. While this is usually ambiguous due to subtractions needed for the
dispersion integrals, the ${\cal N}=4$ theory, which is finite, is special.
There are relations among amplitudes which can be used for this theory.)
If one makes an ansatz for some off-shell extensions
of the tree amplitudes, one can obtain, via unitarity relations, some of the one-loop amplitudes in the nonsupersymmetric theory as well. The off-shell extensions can be checked for consistency in soft-gluon limits, etc., so they are fairly unambiguous.
The results quoted in the beginning of section \ref{sec:7} emerged from such analyses.
Notice that the state of the art here is a combination of rules emerging from
the twistor approach and unitarity relations and a bit of guess work.
The new set of recursion rules also has been very useful \cite{recursion}.

\section{Landau levels and Yang-Mills amplitudes}
\label{sec:9}
\subsection{The general formula for amplitudes}

There is an interesting relationship between the amplitudes of the Yang-Mills theory
and the Landau problem or the problem of quantum Hall effect.
(This is also related to Berkovits' twistor string theory \cite{berkovits}.)
To see how this connection arises, we start by rewriting
the formula (\ref{YM23})  in more compact form as follows.
Define a one-particle wave function for the ${\cal N}=4$ supermultiplet by
\beqar
\Phi (\pi ,\bpi, \eta )&=& \delta \left( {\pi^2 \over \pi^1} - {U^2(\sigma ) \over U^1(\sigma )}\right) \exp \left( {i\over 2} \bpi^{\dot A} \pi^1 {W_{\dot A}(\sigma ) \over U^1(\sigma )}
+i\pi^1 \eta_{\alpha }{\xi^\alpha (\sigma )\over U^1(\sigma  )}\right)\nonumber\\
&=& \delta (\Pi \cdot Z(\sigma )) {Z(\sigma ) \cdot A \over \Pi \cdot A}~
\exp \left( {i\over 2} {\bPi \cdot Z(\sigma )~ \Pi \cdot A \over Z(\sigma )\cdot A}
+i{\Pi\cdot A \over Z(\sigma )\cdot A}\eta\cdot \xi (\sigma )\right)\nonumber\\
\label{LL1}
\eeqar
where we introduced the twistors,
\beq
\Pi^\alpha = (0, \pi^A) = (0, 0, \pi^1, \pi^2), \hskip .4in A_\alpha = (0,0, 1,0)
\label{LL2}
\eeq
which gives $Z\cdot A = U^1$, $\Pi \cdot A = \pi^1$. Notice again that
$\Phi$ is holomorphic in the twistor variables
$(Z, \xi )$ and is invariant under the scaling $Z^\alpha \rightarrow \lambda Z^\alpha$,
$\xi^\alpha \rightarrow \lambda \xi^\alpha$. It is also invariant under the scaling of the
twistor $A_\alpha$. There is also an obvious $SU(4)$ or $SU(2,2)$ invariance, if we
transform $A_\alpha$ as well. Thus the expression for $\Phi$ can be used with
a more general
choice of $A_\alpha$ than the one given in (\ref{LL2}). The twistor $A_\alpha$
plays the role of the reference momentum which has been used in many discussions of
scattering amplitudes; ultimately, it drops out of the physical results due to
conservation of momentum.

On the ${\bf CP}^1$ with the homogeneous coordinates $u^a$, we define
the holomorphic differential
\beq
(udu)\equiv u \cdot du = \epsilon_{ab} u^a du^b
= (u^1)^2 d\sigma
\label{LL3}
\eeq
As a result we can write
\beqar
&&\int d\sigma_1d\sigma_2  \cdots d\sigma_n {1\over (\sigma_1 -\sigma_2)
(\sigma_2 -\sigma_3)\cdots (\sigma_n -\sigma_1)}\nonumber\\
&&\hskip .2in =\int (udu)_1 (udu)_2 \cdots (udu)_n 
{1\over (u_1\cdot u_2) (u_2\cdot u_3)\cdots (u_n\cdot u_1)}
\label{LL4}
\eeqar
Using the formulae (\ref{LL1}, \ref{LL4}), we can write the amplitude
(\ref{YM26}) as
\beqar
{\cal A}
&=& \int d\mu ~ \int \prod_i (udu)_i \Phi (\pi_i, \bpi_i , \eta_i)~
{\Tr (t^{a_1} \cdots t^{a_n})\over (u_1\cdot u_2) (u_2\cdot u_3)\cdots (u_n\cdot u_1)}
\nonumber\\
&&\hskip .3in + {\rm noncyclic ~permutations}
\label{LL5}
\eeqar
Finally, the denominators arise from the chiral Dirac determinant, so we can also write this
as
\beq
{\cal A} =  \int d\mu ~ \int \prod_i (udu)_i \Phi (\pi_i, \bpi_i , \eta_i)~
\left( {\delta \over \delta {\bar{\cal A}}^{a_1}(u_1)}
\cdots {\delta \over \delta {\bar{\cal A}}^{a_n}(u_n)}\right) \Tr \log D_\bz \biggr]_{{\bar{\cal A}}=0}\label{LL6}
\eeq
This formula takes care of the permutations as well. The holomorphic curves of degree $d$ are as given in (\ref{tw20}).

\subsection{A field theory on ${\bf CP}^1$}

We now consider a field theory on ${\bf CP}^1$ or the two-sphere.
The action is given by
\beq
{\cal S} = \int d\mu ({{\bf CP}^1} )~\Bigl[ {\bar q} (\bdel + {\bar{\cal A}} ) q
~+~ {\bar Y} (\bD Q)\Bigr]\label{LL7}
\eeq
Here $q$ and ${\bar q}$ are standard fermionic fields, so that
the first term is the chiral Dirac action on ${\bf CP}^1$.
These fields are analogous to what generates the extra current algebra
in Berkovits' paper \cite{berkovits}.
In the second term, $Q$ stands for the supertwistor variables
$(Z^\alpha, \xi^\alpha)$. ${\bar Y}$ is another field with values in twistor variables again.
Thus the second term corresponds to 
a two-dimensional action with target space ${\bf C}^{4|4}$.
Further, $\bD = \bdel + \bA$, where $\bA$ is a $GL(1, {\bf C})$ gauge field.
This term has invariance under the scaling
$Q \rightarrow \lambda Q$, ${\bar Y} \rightarrow \lambda^{-1} {\bar Y}$,
with the gauge transformation
$\bA \rightarrow \bA  - \bdel \log \lambda$.
Since there are equal number of fermions and bosons, the $GL(1, {\bf C})$ transformation has no anomaly
and the use of the chiral action for the second term is consistent with this symmetry at the quantum level.

We will consider the functional integral of $e^{-{\cal S}}$. The integration over the fermions
$q, {\bar q}$ leads to an effective action $\Tr \log \bD$, and hence, the connected correlator with $n$ factors of ${\bar{\cal A}}$ is given by
\beqar
M&=& \int [d{\bar Y} dZ d\bA ] ~ \exp \left( - \int {\bar Y} \bD Z \right) \times\nonumber\\
&&\hskip .3in \int {d^2\sigma_1 \over \pi}  \cdots {d^2\sigma_n \over \pi}
{\Tr [ {\bar{\cal A}}(1){\bar{\cal A}}(2)\cdots {\bar{\cal A}}(n)]\over
(\sigma_1-\sigma_2)(\sigma_2-\sigma_3)  \cdots (\sigma_n - \sigma_1)}\label{LL8}
\eeqar
We will now take ${\bar{\cal A}}$ to be of the form ${\bar{\cal A}} = \bdel \Phi$.
We can use this in (\ref{LL8}) and carry out the integration over
$d{\bar \sigma}$'s. Partial integration can produce $\delta$-functions like
$\delta^{(2)}(z_1 -z_2)$. If we exclude coincidence of points,
which will correspond to coincidence of external momenta
in the context of multigluon scattering, then these $\delta$-functions have no support and the only contribution is from the boundary. The correlator (\ref{LL8})  then becomes
\beqar
M&=& \int [d{\bar Y} dZ d\bA ] ~ \exp \left( - \int {\bar Y} \bD Z \right) \times\nonumber\\
&&\hskip .3in \oint {d\sigma_1 \over 2\pi i}  \cdots {d\sigma_n \over 2\pi i}
{\Tr [\Phi (1) \Phi (2)\cdots \Phi (n)]\over
(\sigma_1-\sigma_2)(\sigma_2-\sigma_3)  \cdots (\sigma_n - \sigma_1)}\label{LL9}
\eeqar
Using the result (\ref{LL4}), we can rewrite this as
\beqar
M&=& \int [d{\bar Y} dZ d\bA ] ~ \exp \left( - \int {\bar Y} \bD Z \right) \times\nonumber\\
&&\hskip .3in \oint {(u\cdot du)_1 \over 2\pi i}  \cdots {(u\cdot du)_n \over 2 \pi i}
{\Tr [\Phi (1) \Phi (2)\cdots \Phi (n)]\over
(u_1\cdot u_2)(u_2\cdot u_3)\cdots (u_n\cdot u_1)  }\nonumber\\
\label{LL9}
\eeqar

We now turn to the integration over the gauge field $\bA$. There are nontrivial $U(1)$ bundles over $S^2 \sim {\bf CP}^1$, corresponding to Dirac monopoles. The same result holds for $GL(1, {\bf C})$.
The space of gauge potentials then splits up into a set of disconnected pieces, one for each monopole number. In each sector, we can write
$\bA = \bA_d + \delta \bA$, where $\bA_d$ is a fixed configuration of monopole number
$d$ and $\delta \bA$ is a fluctuation (of zero monopole number).
In two dimensions, we can always write $\delta \bA = \bdel \Theta$ for some complex
function $\Theta$ on ${\bf CP}^1$.The measure of integration splits up as
$[d\bA ] = [d\Theta ]~\det \bdel $. The determinant can contribute to the conformal anomaly, if we interpret this as the world-sheet formulation of a string theory calculation.
For us, thinking of this as a two-dimensional field theory, this anomaly is not relevant.
We can now eliminate $\Theta$ by absorbing it into the definition of the fields.
$\Phi$ will be chosen to be $GL(1, {\bf C})$ gauge-invariant, so this will not affect it.
The integration over the fields $\bA$ is thus reduced to a summation over
the different monopole numbers, with a fixed representative background field $\bA_d$ for each
value of $d$.

For the integration over the twistor fields, we need a mode expansion.
For the sector with monopole number $d$, we can expand the fields as
\beqar
Z^\alpha &=& \sum_{\{a\}} a^\alpha_{a_1 a_2 \cdots a_d} ~ u^{a_1} u^{ a_2} \cdots u^{ a_d}
 ~+ {\rm higher~Landau ~levels}\nonumber\\
\xi^\alpha &=& \sum_{\{ a\}} \gamma^\alpha_{a_1 a_2 \cdots a_d} ~ u^{a_1} u^{ a_2} \cdots u^{ a_d} ~+ {\rm higher~Landau ~levels}\label{LL10}
\eeqar
with similar expansions for ${\bar Y}$. The first set of terms correspond to 
the lowest Landau level, the higher terms, which we have not displayed
explicitly,  correspond to higher Landau levels. The wave functions (or mode functions) for the lowest Landau level are holomorphic in the $u$'s, the higher levels involve ${\bar u}$'s as well.
The functional integration is now over the coefficients
$a^\alpha_{a_1 a_2 \cdots a_d} $, $\gamma^\alpha_{a_1 a_2 \cdots a_d}$, etc.
Notice that the zero modes define a holomorphic curve of degree $d$
in supertwistor space. Our choice of $\Phi$ will have no ${\bar Y}$, so, in integrating
over the nonzero modes, one cannot have any propagators or loops generated by 
${\bar Y}-Z$ Wick contractions. As a result, the nonzero modes give only an overall
normalization factor. In any correlator involving only $Z$'s and $\xi$'s, we can saturate
the fields by the zero modes. Then, apart from constant factors, the correlator 
(\ref{LL9}) becomes
\beqar
M&=& \sum_d C_d {\cal M}_d\nonumber\\
{\cal M}_d &=& \int d\mu (a, \gamma ) \oint {(u\cdot du)_1 \over 2\pi i}  \cdots {(u\cdot du)_n \over 2 \pi i}
{\Tr [\Phi (1) \Phi (2)\cdots \Phi (n)]\over
(u_1\cdot u_2)(u_2\cdot u_3)\cdots (u_n\cdot u_1)  }\nonumber\\
\label{LL11}
\eeqar

We now choose the $\Phi$'s as given in (\ref{LL1}), and the correlator (\ref{LL11}) becomes the multigluon amplitude given in (\ref{LL6}), once we can show that the integration over the moduli space is given by the invariant measure (\ref{YM28}). To see how this arises,
consider an $SL(2, {\bf C})$ transformation of the coefficients of the mode expansion given by
\beqar
a'^\alpha_{a_1 a_2 \cdots a_d} &=& a^\alpha_{b_1 b_2 \cdots b_d} g^{b_1}_{a_1}
 g^{b_2}_{a_2}\cdots  g^{b_n}_{a_n}\nonumber\\
\gamma'^\alpha_{a_1 a_2 \cdots a_d}&=& \gamma^\alpha_{a_1 a_2 \cdots a_d}
g^{b_1}_{a_1}
 g^{b_2}_{a_2}\cdots  g^{b_n}_{a_n}\label{LL12}
\eeqar
where $g \in SL(2, {\bf C})$. If we use $( a', \gamma' )$, this is equivalent to
using $(a, \gamma )$ and redefined $u$'s, $u'^a = g^a _b u^b$
in $\Phi$. Since $u\cdot du$ and scalar products like
$(u_1\cdot u_2)$ are invariant under such transformations, we can change the variables
$u' \rightarrow u$; the integrand for the integration over the moduli
$(a, \gamma )$ is thus $SL(2, {\bf C})$-invariant. We must therefore consider the
measure (\ref{YM28}) to obtain well-defined correlators.

What we have shown is that there is a set of correlators in the two-dimensional problem
defined by (\ref{LL7}) which can be calculated exactly, being saturated by the lowest Landau levels, and which give the multigluon amplitudes.
The YM amplitudes are thus obtained as a set of ``holomorphic'' correlators
of the two-dimensional problem. The fermionic integration obviously leads to an expression which is similar to the Laughlin wave function for the $\nu =1$ quantum Hall state.
If we consider just the value $\alpha =1$, the fermionic integration has the form
\beq
\int [d\gamma ] \prod_i \gamma^1_{a_1 a_2 \cdots a_n} u_i^{a_1} \cdots u_i^{a_n}
\sim \prod_{i<j} (u_i\cdot u_j )
\label{LL13}
\eeq
(With four sets of such terms for $\alpha =1$ to $\alpha =4$, we get the
fourth power of the right hand side.) In some sense, we can interpret the integration over the
bosonic moduli as defining the bosonic version of the Laughlin wave function, since it is
related to the fermionic one by the natural supersymmetry of the expression $\Phi$
in (\ref{LL1}). Whether this Landau level point of view for the amplitudes
can lead to new insights
into the YM problem is yet to be seen.
\vskip .2in
This work
was supported by the National Science Foundation grant number
PHY-0244873 and by a PSC-CUNY grant.

\end{document}